\documentclass[ acmtog, %
                authorversion   =true,
                screen          =true,
                authordraft     =false,
                review          =false,
                timestamp       =false,
                balance         =true,
                anonymous       =false]{acmart}

\AtBeginDocument{%
  }

\copyrightyear{2022}
\acmYear{2022}
\setcopyright{rightsretained}
\acmConference[SIGGRAPH Asia '22 Conference Papers]{SIGGRAPH Asia 2022 Conference Papers}{December 6--9, 2022}{Daegu, Republic of Korea}
\acmBooktitle{SIGGRAPH Asia 2022 Conference Papers (SA '22 Conference Papers), December 6--9, 2022, Daegu, Republic of Korea}\acmDOI{10.1145/3550469.3555425}
\acmISBN{978-1-4503-9470-3/22/12}

\usepackage{flushend}
\usepackage{caption}
\usepackage{subcaption}
\usepackage{comment}
\usepackage{array}
\usepackage{xspace}
\usepackage{breakcites}
\usepackage{siunitx}
\usepackage{xurl}

\DeclareMathOperator{\softmax}{softmax}
\DeclareMathOperator{\softmin}{softmin}

\newcommand{\parent}[1]{\left( #1 \right)}

\newcommand{\braces}[1]{\left\{ #1 \right\}}
\newcommand{\emphasis}[1]{\emph{#1}}
\newcommand{\makename}[3][s]{%
  \expandafter\newcommand\csname #2\endcsname{#3}%
  \expandafter\newcommand\csname #2s\endcsname{#3#1}%
} %

\makename{Scene}{Layout}
\makename{scene}{layout}
\makename{Furnobj}{Furniture object}
\makename{furnobj}{furniture object}
\makename{Score}{Loss}
\makename{score}{loss}

\newcommand{\furn}[2][]{F^{#1}_{#2}} %
\newcommand{\cat}[2][]{c^{#1}_{#2}} %
\newcommand{\ori}[2][]{o^{#1}_{#2}} %
\newcommand{\width}[2][]{w^{#1}_{#2}} %
\newcommand{\depth}[2][]{d^{#1}_{#2}} %
\newcommand{\xpos}[2][]{x^{#1}_{#2}} %
\newcommand{\ypos}[2][]{y^{#1}_{#2}} %
\newcommand{\sample}[1]{S^{#1}} %
\newcommand{\loss}[1]{E_{#1}} %
\newcommand{\norm}[1]{\left\lVert#1\right\rVert}

\newcommand{\methodname}{LayoutEnhancer\xspace}

\graphicspath{{figures/}} %

\begin{document}

\title[\methodname: Generating Good Indoor Layouts from Imperfect Data]{\methodname: Generating Good Indoor Layouts from Imperfect Data}

\author{Kurt Leimer}
\affiliation{
  \institution{NJIT/TU Wien}
   \country{United States/Austria}
}
\author{Paul Guerrero}
\affiliation{
  \institution{Adobe Research}
   \country{United Kingdom}
}
\author{Tomer Weiss}
\affiliation{
    \institution{NJIT}
    \country{United States}
}
\author{Przemyslaw Musialski}
\affiliation{
    \institution{NJIT}
    \country{United States}
}

\renewcommand{\shortauthors}{Kurt Leimer, Paul Guerrero, Tomer Weiss, Przemyslaw Musialski }

\begin{abstract}
We address the problem of indoor layout synthesis, which is a topic of continuing research interest in computer graphics. 
The newest works made significant progress using data-driven generative methods; however, these approaches rely on suitable datasets. In practice, desirable layout properties may not exist in a dataset, for instance, specific expert knowledge can be missing in the data. 
We propose a method that combines expert knowledge, for example, knowledge about ergonomics, with a data-driven generator based on the popular Transformer architecture. The knowledge is given as differentiable scalar functions, which can be used both as weights or as additional terms in the loss function. 
Using this knowledge, the synthesized layouts can be biased to exhibit desirable properties, even if these properties are not present in the dataset. Our approach can also alleviate problems of lack of data and imperfections in the data. 
Our work aims to improve generative machine learning for modeling and provide novel tools for designers and amateurs for the problem of interior layout creation. 
\end{abstract}

\ccsdesc[500]{Computing methodologies~Computer graphics}
\ccsdesc[500]{Computing methodologies~Neural networks}
\ccsdesc[300]{Computing methodologies~Computer vision}

\keywords{neural networks, layout synthesis, interior design, interior layout generation}

\begin{teaserfigure}
\includegraphics[width=\textwidth]{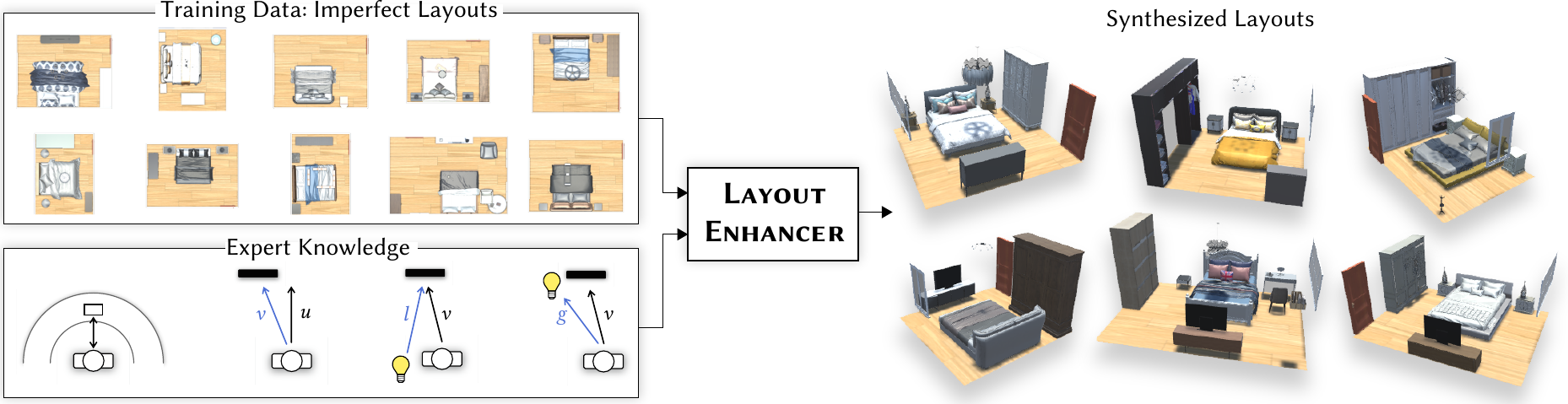}\\
\caption{Our proposed \methodname combines data-driven learning from potentially imperfect data with expert knowledge. Generated layouts are biased to follow rules laid out in the expert knowledge, effectively reducing the impact of data imperfections. See Figure~\ref{fig:bad_example} for examples of imperfections that are avoided due to the inclusion of expert knowledge.
}
\label{fig:teaser}
\end{teaserfigure}

\maketitle

\section{Introduction}
\label{sec:intro}

Indoor spaces play a central role in our everyday lives.
The synthesis and design of indoor layouts (apartment layout, workplace layout) is a long-standing problem in several disciplines, including graphics~\cite{merrell2011interactive,fisher2012example}. %

In this paper, we address the problem of data-driven layout synthesis that has recently gained renewed interest in computer graphics due to the advent of a novel neural methods in generative machine learning~\cite{paschalidou2021atiss, para2021generative}.  
However, despite recent progress, interior layout synthesis is still challenging for machine learning algorithms. The problem is twofold: 

First, reliable training data is difficult to obtain. Designs need to be crafted manually by professionals, making the process labor- and time-intensive and hence expensive.

\begin{figure}[t!]
    \centering
    \includegraphics[width=0.999\columnwidth]{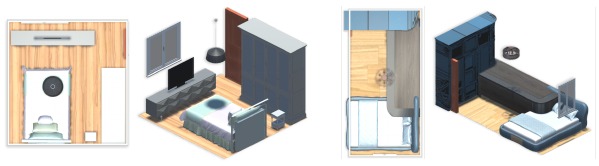}
    \caption{LayoutEnhancer can learn to
    improve issues found in imperfect data like \emph{ergonomic issues} (left room): (i) a window directly behind the TV causes glare on sunny days, making it difficult to watch due to a big contrast in brightness. (ii) Insufficient illumination for reading a book without a light source behind or beside the bed; and \emph{geometric issues} (right room): (i) desk is intersecting with the bed and the closet; (ii) closet is covering the door.}
    \label{fig:bad_example}
\end{figure}

Second, readily available datasets may have been created by non-experts and may contain several issues like incorrect intersections, unrealistic placement, misplaced objects, etc. (cf. Figure~\ref{fig:bad_example}). 
At the same time, high-quality indoor design requires expert knowledge because good furniture arrangements are connected to several considerations like functionality, usability, aesthetics, cost-effectiveness, and ergonomics. These may not all be reflected in a dataset, which contains layouts that were most likely not created by interior design experts.

We address these problems by using a Transformer-based generative model with additional expert knowledge ``injected'' into the data-driven training process.
Transformers are generative models originally proposed for natural language processing that have proven very successful in a wide range of domains~\cite{vaswani2017attention}. Recently, several methods have successfully used transformers for layout generation~\cite{para2021generative,wang2020sceneformer,paschalidou2021atiss}.

In our approach, a layout $S$ is defined as a sequence of discrete elements $S \coloneqq \{F_0, \dots, F_N\}$, each represented with a fixed-length parameter vector. A traditional generative model learns to generate new layouts according to a probability distribution $p(S)$ that approximates the probability distribution of the dataset $p(S) \approx p_{\text{data}}(S)$.

We propose to inject additional information based on expert knowledge into the learning process to obtain a learned distribution $p'(S)$
that reflects both the dataset distribution and the additional information.
The expert knowledge biases the learned probability distributions to emphasize or de-emphasize specific properties of the layouts.  
In Section~\ref{sec:scores} we derive a set ergonomic rules from expert literature~\cite{kroemer2017fitting}. %

We integrate this information into the loss function of our trans\-former-based generative model in two ways: (i) as weights of training samples and (ii) as additional loss that assesses the quality of samples proposed during the training process. In the second case, expert knowledge needs to be differentiable w.r.t. the predicted probabilities. 
We discuss the details in Section~\ref{sec:gen}.

In Section~\ref{sec:results}, we evaluate the proposed method and compare it to a recent data-driven method that does not utilize expert knowledge~\cite{paschalidou2021atiss}. We demonstrate that with our approach we can improve the ergonomic quality of generated layouts, effectively increasing the perceived realism compared to others.

In summary, the contributions of this paper are three-fold: 
\begin{itemize}
\item 
We introduce a differentiable ergonomic loss that can be used to assess the ergonomic quality of interior layouts. We derive this loss from the expert knowledge in ergonomics (Section~\ref{sec:scores}). 
\item 
We integrate this differentiable loss into the training of a Transformer network (Section~\ref{sec:gen}). 
\item 
We empirically show that we can train a generative model with this loss that creates samples with increased ergonomic quality and realism compared to the state of the art
(Section~\ref{sec:results}). 
\end{itemize}

\section{Related Work}\label{sec:related}

Interior spaces and their layouts are part of everyday life.
For example, organizations such as Ikea and Wayfair are actively working toward understanding their customers needs~\cite{ataer2019room}.
Typically, each domain has different requirements and needs, which require manual design~\cite{wayfairroomplanner}.

In practice, designing layouts is a laborious task due to high dimensional design space, ranging from selecting relevant furniture pieces, to arranging the target space to fit the design goals.
To alleviate such manual workflow, researchers have proposed multiple computational methods to assist in layout design. 
Below we classify previous work based on their approach.

\paragraph{Deep Learning Methods}
Such methods employ neural networks, in which the network learns layout patterns from images, graphs, or other data.
Such 3d scene data and the data modality is an important factor in deep learning~\cite{fu2021front}.
Early deep learning work utilizes top-down images of layouts to understand object-object layout relationships~\cite{wang2018deep}.
However, images do not naturally contain sufficient detail for the 
network to synthesize complex human-centered layouts. 
Graphs have also been proposed as a means to encode spatial layout information~\cite{wang2019planit,zhou2019scenegraphnet,Luo2020}. 

In addition to images and graphs, researchers explored how to use other 3d scene data representations for synthesis. 
Zhang et al.~\shortcite{li2019grains} synthesize scenes by sampling from a vector that represents the spatial structure of a scene. 
Such structure encodes a hierarchy of geometrical and co-occurrence relations of layout objects. 
\cite{zhang2020deep} proposed a hybrid approach that combines such vector representation with an image-based approach. Also other utilize graph structures to describe scene layouts~\cite{Di2020}. 
Yang et al.~\shortcite{yang2021scene} combine such vector representation with Bayesian optimization to improve 
furniture placement predictions of the generative network. Recently, variational autoencoders have been proposed for indoor layout synthesis~\cite{Chattopadhyay2022StructuredGV}. 

Most recently, researchers have proposed to use neural networks based on transformers~\cite{wang2020sceneformer,paschalidou2021atiss}. 
However, in contrast to our method, their work does not account for ergonomic qualities which results in misplaced furniture items.

\paragraph{Other Approaches}
Before the era of deep learning, early work considered layout synthesis as a mathematical 
optimization problem, where a set of constraints describe the layout quality in terms of energy an energy functional~\cite{yu2011make,merrell2011interactive,weiss2018fast}. The layout is then optimized via stochastic or deterministic optimization process. 

Other researchers proposed data-driven methods.
Qi et al.~\shortcite{qi2018human} use interaction affordance maps for each layout object for stochastic layout synthesis. 
Similarly, Fisher et al.~\shortcite{fisher2015activity} used annotated 3d scans of rooms to identify which activities does an environment support. 
Other researchers also learn layout structure from 3d scans for scene synthesis~\cite{kermani2016learning}. 
They extract manually defined geometric relationships between objects from such scans, which are then placed using a stochastic optimization. 

Other research has made progress towards incorporating human-centered considerations for 3d scene synthesis. 
Fu et al.~\shortcite{fu2017adaptive} use a graph of objects to guide a layout synthesis process. 
However, they only consider static human poses in relation to activities.
Zhang et al.~\shortcite{zhang2021joint} and Liang et al.~\shortcite{liang2019functional} focus on optimal work-space design. 
While the authors demonstrate novel use of simulation and dynamic capture of agent in action metrics, they only focus on mobility and accessibility based factors. 
In \cite{puig2018virthome}, the authors demonstrate how to evaluate the functionality of layouts.
However, this work does not include 3d scene synthesis. 

\begin{figure}[t]
    \centering
    \includegraphics[width=1.0\columnwidth]{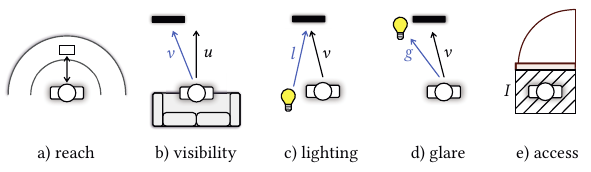} %
    \caption{Ergonomic rules implemented in our system. We chose these guidelines as they are essential in most indoor scenarios, like reading a book, watching TV, or working at the desk or the computer. We convert the rules to scalar cost functions and evaluate them using activities (cf. Section~\ref{sec:scores}). %
    }
    \label{fig:scores}
\end{figure}

Early work ~\cite{yu2011make,merrell2011interactive} has also included ergonomic and interior design knowledge into the layout design process. Our approach differs from these existing methods in two major aspects. First of all, their methods require the manual definition of a number of additional layout design rules. Second, their methods are designed to optimize the arrangement of an existing furniture layout, while our approach can synthesize entirely new layouts with desired characteristics.

\section{Ergonomic Costs}\label{sec:scores}

To derive a set of rules used to quantify an ergonomic quality of a design, we studied the literature of ergonomic guidelines~\cite{kroemer2017fitting}. As a result, we order the information in a hierarchical manner, using the building blocks of activities, actions and ergonomic costs. 

An activity is a set or sequence of actions that need to be performed to accomplish a specific goal \cite{puig2018virthome}. An activity could be, for instance, reading a book or watching TV. A single action puts specific elements of a layout into a common context, for example looking at the TV while sitting on the sofa. Ergonomic costs are evaluated for each action to quantify how suitable the arrangement of the layout elements is in an ergonomic sense. 

The ergonomic \scores\ obtained for each evaluated ergonomic rule are then aggregated up the hierarchy to obtain the \scores\ for each action, activity and finally for the whole layout. This formulation makes it easy to define new evaluation functions for different activities by combining the various building blocks. 
In our approach, we consider the following ergonomic costs (cf. Figure~\ref{fig:scores}):
\begin{itemize}
	\item Reach measures how easy it is to interact with a target object from a given position.
 	\item Visibility measures how visible a target object is for a given position and viewing direction.
 	\item Lighting measures how well an object is illuminated by light sources in the room.
 	\item Glare measures the decrease in visual performance from strong brightness contrast caused by having bright light sources in the field of view.
 	\item Accessibility measures how much free space is in front of a target object to allow easy interaction and walking by.
\end{itemize}
We choose the above five rules as examples for two reasons. First, they are all relevant for the kinds of activities that are often performed in the prevalent room types that are included in publicly available indoor layout datasets . The second reason is a practical one, since these rules can be defined as (piecewise) differentiable scalar functions in a range of $[0,1]$, which perfectly suits our needs.

\begin{figure}
    \centering
    \includegraphics[width=0.47\textwidth]{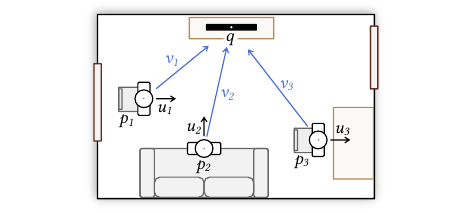} %
    \caption{Human activity in the room based on the example of \emph{Watch TV}. For all possible sitting locations $p_j$ an avatar is sampled and the ergonomic rules for visibility and glare are evaluated. The final contribution is the weighted sum of costs over every combination of a sitting possibility $p_j$ and all TVs $q_k$. Please refer to Section~\ref{sec:scores} for more details.}
    \label{fig:activity}
\end{figure}

For instance, given a target object at position $q_k$ viewed from position $p_j$ and viewing direction $u_j$, we define the \textit{visibility cost} as smooth scalar function $E_V$ of two vectors $u_j$ and $v = \frac{q_k-p_j}{\norm{q_k-p_j}}$ which can be minimized:
\begin{equation*}
	\loss{V} = 1 - \left(\frac{ 1 + \langle u_j, v \rangle }{2} \right) \,.
\end{equation*}

Together with the \textit{glare cost} function $E_G \left(p_j,B,q_k \right)$ with light sources $B$, we can compute the \score\ for the activity \emph{Watch TV} (cf.  Figure~\ref{fig:activity}):
\begin{equation*}
	e^{tv}_{j,k} = \frac{E_V \left(p_j,u_j,q_k \right) + E_G \left(p_j,B,q_k \right)}{2} \,.
\end{equation*}
Since there can be multiple TVs in a room in addition to multiple pieces of seating furniture, we need to compute the weighted sum of costs over every combination of $p_j$ and $q_k$, using $e^{tv} = [e^{tv}_{j,k}]_{j \in P,k \in Q}$:
\begin{equation*}
	E_{tv} = \langle e^{tv}, \softmin(\beta \cdot e^{tv}) \rangle \,.
\end{equation*}
The costs of every possible activity are then aggregated to obtain the total \textbf{ergonomic loss} $E$ of the \scene. Figure~\ref{fig:scores} depicts all five ergonomic cost functions implemented in our framework in a similar differentiable fashion. We refer the reader to supplemental material for details on the implementation of the other ergonomic cost functions and activities.

\section{\Scene\ Generation with Expert Knowledge}
\label{sec:gen}

We build on top of Transformers~\cite{vaswani2017attention} as generative model for layouts~\cite{para2021generative, wang2020sceneformer, paschalidou2021atiss}. 
In this section, we first present our model and then describe how we integrate our ergonomic \score\ into the  training.

\subsection{Generative Model}

Transformers are sequence generators that originate from natural language processing. A \scene\ is generated step-wise as a sequence of discrete tokens $S=(s_1, \dots, s_n)$, one token $s_i$ at a time. Thus, we first need to define a sequence representation of our \scene s. 

\begin{figure}
    \centering
    \includegraphics[width=0.488\textwidth,trim={0 23 0 0},clip]{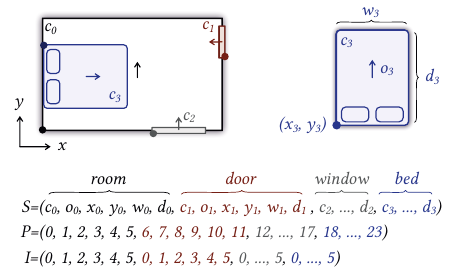} %
    \caption{A \scene\ is represented as a sequence $S=(s_1, \dots, s_n)$. Each individual token $s_i$ in the sequence represents an attribute of a \furnobj, such as its category, orientation, position or dimensions. %
    }
    \label{fig:sequences}
\end{figure}

\paragraph{Sequence representation}

Each \furnobj\ is represented as a 6-tuple $\furn{i} = (\cat{i},\ori{i},\xpos{i},\ypos{i},\width{i},\depth{i})$, with $\cat{i}$ indicating the object category, such as \emph{chair} or \emph{table}, $\ori{i}$ the orientation, $\xpos{i}$ and $\ypos{i}$ being the x- and y-coordinates of the bottom left corner of the \furnobj, $\width{i}$ being the width, and $\depth{i}$ the depth  of the \furnobj\ (cf.~Figure~\ref{fig:sequences}) . Since previous work~\cite{paschalidou2021atiss} has shown that randomizing the order of objects that do not admit a consistent ordering can be beneficial, we follow a similar approach. The bounding box of the room itself is represented as the \furnobj\ $\furn{0}$ and is thus always the first of the ordered \furnobjs, followed by the doors and windows of the layout. The order of all other \furnobjs\ is not consistent and instead randomized during training. We concatenate the 6-tuples of the ordered \furnobj s and add a special stop token to the end of the sequence to obtain the sequence $S$. An example can be seen in Figure \ref{fig:sequences}.

Similar to previous work~\cite{wang2020sceneformer}, we use two additional parallel sequences to provide context for each token in $S$: a position sequence $S^P = (1, 2, \dots, n)$ that provides the global position in the sequence, and an index sequence $S^I = (1, 2, \dots, 6, 1, 2 \dots, 6)$ that describes the index of a token inside the 6-tuple of a \furnobj.

Our approach also supports an alternate method of providing the room shape as a binary map of the floor plan, similar to ATISS~\shortcite{paschalidou2021atiss}. While specifying the room as part of the sequence allows the network to learn how to synthesize arbitrary rectangular rooms, using a binary map instead lets the network learn how to generate furniture layouts for more complex non-rectangular room shapes. %

\paragraph{Quantization} 
Transformers typically operate with discrete token values. By learning to predict a probability for each possible value of a token, a transformer can model arbitrary distributions over token values. To obtain discrete values, we quantize all object parameters except orientations $o_i$ and categories $c_i$ uniformly between the minimum and maximum values that occur in the dataset. Orientations $o_i$ are uniformly quantized in $[0, 2\pi)$, adjusting the resolution to preserve axis-aligned orientations as integer values. We use a resolution of $r = 256$. Categories $c_i$ do not require quantization as they are already integers. We use categorical distributions for all tokens.

\begin{figure}
    \centering
    \includegraphics[width=0.46\textwidth]{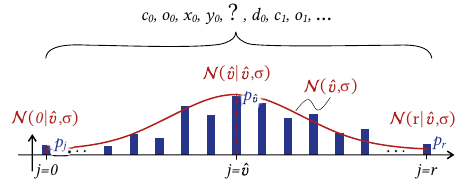} %
    \caption{To propagate the ergonomic loss back to the token probabilities, we choose the maximum of the discrete values of the predicted token and convolve the neighborhood with a Gaussian kernel, centered at the discrete maximum. The resulting token value is a weighted sum of the discrete values in this neighborhood, weighted by the probability and distance to the kernel center of each discrete value. Please refer to Section \ref{sec:loss_propagation} for more details. }
    \label{fig:interpolation}
\end{figure}

\paragraph{Sequence generation} 
Our Transformer-based sequence generator $f_\theta$ factors the probability distribution over sequences $S$ into a product of conditional probabilities over individual tokens:
\begin{equation*}
    p(S| \theta) = \prod_i p(s_i | s_{<i}, \theta),
\end{equation*}
where $s_{<i} \coloneqq s_1, \dots, s_{i-1}$ is the partial sequence up to (excluding) $i$. Given a partial sequence $s_{<i}$, our model predicts the probability distribution over all possible discrete values for the next token: $p(s_i | s_{<i}, \theta) = f_\theta(s_{<i}, s_{<i}^{P}, s^I_{<i})$ that can be sampled to obtain the next token $s_i$. Here $s_{<i}^{P}$ and $s^I_{<i}$ are the corresponding partial position and index sequences that are fully defined by the index $i$.
We implement $f_\theta$ as a GPT-2 model~\cite{radford2019language} using the implementation included in the Huggingface library \cite{wolf2020transformers}.

\subsection{Ergonomic Loss}\label{sec:loss_propagation}

\begin{figure*}
    \centering
    \includegraphics[width=0.98\textwidth]{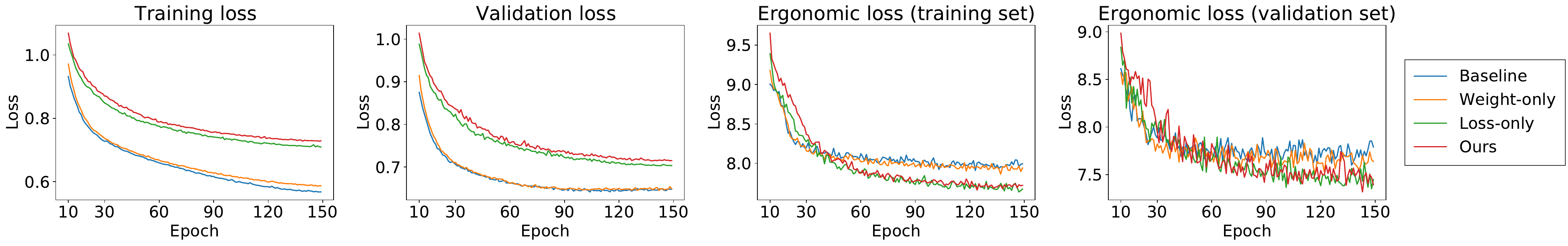}
    \caption{Cross-entropy loss and ergonomic loss for our model and its ablations, evaluated on the Bedrooms dataset. The training loss and validation loss refer to the cross-entropy loss on the training and validation sets, respectively. By including our proposed ergonomic loss term during training we can significantly decrease the ergonomic loss of synthesized \scenes.
    }
    \label{fig:plot_bedrooms}
\end{figure*}

A loss designed by an expert, such as an ergonomic rule, defines desirable properties of \scene s that may not be fully realized in a dataset. 
However, while minimizing the expert loss may be \emph{necessary} to obtain a desirable \scene, it is usually not \emph{sufficient}, since a manually defined loss can usually not describe \emph{all} desirable properties of a \scene\ exhaustively.
Our goal is thus to combine the expert loss with a data-driven generative model for \scene s. 
However, integrating the ergonomic \score\ in a transformer-based generative model poses two main challenges: 

\textbf{C1}: Transformers generate \scene s in multiple steps, each step generating a small part of the \scene\ such as a single object or a single object attribute. Each step, where only a partial \scene\ has been generated, requires supervision, but the ergonomic \score\ cannot reliably be computed on a partial \scene. 

\textbf{C2}: The ergonomic \score\ is defined over continuous parameters, such as object positions or orientations. However, transformers typically output a probability distribution over a discrete set of values in each step, such as quantized object positions or orientations. This makes gradient propagation from the ergonomic loss to the transformer difficult.

To tackle the first challenge (\textbf{C1}), we observe that transformers are typically trained with a strategy called \emph{teacher forcing}, where the partial sequence $s_{<i}$ preceding the current token $s_i$ is taken from a ground truth \scene. Thus, when generating a token $s_i$, we can evaluate the ergonomic \score\ on the \scene\ defined by $s_{<i}, s_i, s_{>i}$, where only $s_i$ is generated and both the preceding tokens $s_{<i}$ and the following tokens $s_{>i}$ are taken from the ground truth, effectively evaluating $s_i$ in the context of the ground truth \scene.

To solve the second challenge (\textbf{C2}) 
we need an ergonomic loss that is differentiable w.r.t. the probabilities $p(s_i | s_{<i}, \theta)$ predicted by our generative model. A straight-forward solution computes the expected value of the ergonomic \score\ $E$ over all possible values $v_j$ of a token $\sum_j E(s_{<i},\ v_j,\ s_{>i}) P(s_i=v_j | s_{<i}, \theta)$. This solution is differentiable w.r.t. the probabilities, but requires an evaluation of the ergonomic \score\ for each possible value of a token, which is prohibitively expensive. Instead, we opt for a less exact but much more efficient approach, where only a single evaluation of the ergonomic \score\ per token is needed. We compute the ergonomic loss $\mathcal{L}_E$ as the ergonomic \score\ for the expected value of a token in a small window around the most likely value of the token:
\begin{gather}
	\mathcal{L}_E = E(s_{<i},\ \bar{v},\ s_{>i}), \text{ with}\\
	\bar{v} = \frac{\sum_{j} \left( \mathcal{N}(v_j | \hat{v}, \sigma)\ P(s_i=v_j | s_{<i}, \theta)\ v_j \right)}{\sum_{j} \left( \mathcal{N}(v_j | \hat{v}, \sigma)\  P(s_i=v_j | s_{<i}, \theta) \right)}, \nonumber
\end{gather}
where $\mathcal{N}(x|\hat{v}, \sigma)$ is the normal distribution centered at $\hat{v}$ with standard deviation $\sigma$. $\hat{v}$ is the token value with highest probability, and $\sigma$ is set to $1/r$ in our experiments. Figure \ref{fig:interpolation} illustrates the approach. This loss provides gradients to all values in smooth window. Note that increasing the size of the window by increasing $\sigma$ would propagate the gradient to a larger range of token values, but could also result in expected token values $\bar{v}$ that are in low-probability regions of the distribution $p(s_i | s_{<i}, \theta)$, since the distribution may be multi-modal.
The total loss function $\mathcal{L}$ is then given by
\begin{equation}\label{eq:total_loss}
\begin{aligned}
    \mathcal{L} \parent{\sample{k}} = \beta_T \mathcal{L}_T \parent{\sample{k}} + \beta_E \mathcal{L}_E \parent{\sample{k}},
\end{aligned}
\end{equation}
with $\mathcal{L}_T$ being the cross-entropy loss, $\mathcal{L}_E$ being our proposed ergonomic loss and $\beta_T, \beta_E$ being weights that determine the influence of the two loss terms to the overall loss. We use $\beta_T = 1-E\parent{\sample{k}}$ and $\beta_E = E\parent{\sample{k}}$, such that the cross-entropy loss has higher influence for training samples with better ergonomic \score\, while the ergonomic loss is more important for samples with lower ergonomic \score. Essentially, we want the network to learn about the general target distribution from examples that are already considered good, while learning how to improve the ergonomic \score\ from bad examples. %
In Section \ref{sec:res:ablation}, we discuss the influence of the weights $\beta_T$ and $\beta_E$ in more detail.

\subsection{Training and Inference}

We train our models using the 3DFRONT dataset \cite{fu2021future,fu2021front} as training data. 
During training, we randomly augment each training sample by horizontal mirroring and/or rotation in $90^{\circ}$ steps, in addition to applying a random permutation on the order of \furnobjs\ other than the room, windows and doors. For inference, we follow a similar approach to the strategy proposed by Sceneformer \cite{wang2020sceneformer}, using top-p nucleus sampling with $p=0.9$ for the object categories, as well as the attributes of the room, doors and windows. For the attributes of other object categories, we always pick the token with the highest probability. We also check for intersections after sampling each \furnobj\ and re-sample the current object if it cannot be inserted into the layout without intersecting other objects.

\section{Results and Evaluation}
\label{sec:results}

\subsection{Ablation}\label{sec:res:ablation}

To evaluate the influence of our proposed ergonomic loss, we define $3$ ablations of our network that are trained with different loss functions. Recall that the total loss function of our approach given in Eq. \ref{eq:total_loss} is defined as the weighted sum of the cross-entropy loss $\mathcal{L}_T$ and the ergonomic loss $\mathcal{L}_E$ with weights $\beta_T, \beta_E$. Using these weight parameters, we define the following $3$ ablations of our network:
\begin{itemize}
	\item Baseline, with $\beta_T = 1$ and $\beta_E = 0$,
	\item Weight-only, with $\beta_T = 1-E\parent{\sample{k}}$ and $\beta_E = 0$,
	\item Loss-only, with $\beta_T = 1$ and $\beta_E = 1$.
\end{itemize}
In other words, the baseline model only uses the cross-entropy loss with each input sample having equal weight and is thus without any of our enhancements. The weight-only model uses the cross-entropy loss with each sample being weighted by its ergonomic \score, while the loss-only model uses the sum of cross-entropy loss and ergonomic loss with each input sample having equal weight.

Figure \ref{fig:plot_bedrooms} depicts the cross-entropy loss and ergonomic loss evaluated on both the training and validation sets for each version, using the Bedroom dataset for training. The results show a decrease in ergonomic loss for both the loss-only model and our full model which make use of our ergonomic loss term during training. While the decrease may seem small relative to the overall loss, please keep in mind that the loss is computed for the entire scene with only one token predicted by the network. The weight-only model only yields a small decrease of ergonomic loss during training, since weighting the training samples by their ergonomic \score\ only reduces the influence of bad training samples without teaching the network how to improve the sample. However, this still has a noticeable effect on the synthesized scenes as we will discuss in Section \ref{sub:user_study}. Please note that our loss-only model and our full model exhibit a higher cross-entropy loss for both training and validation set. This result is expected, since we aim to improve the ergonomic qualities of the synthesized \scenes\ instead of perfectly recreating the distribution of the dataset.

\subsection{Room-conditioned Layout Synthesis}\label{sub:user_study}

\begin{figure}[t]
    \centering
    \includegraphics[width=0.98\columnwidth]{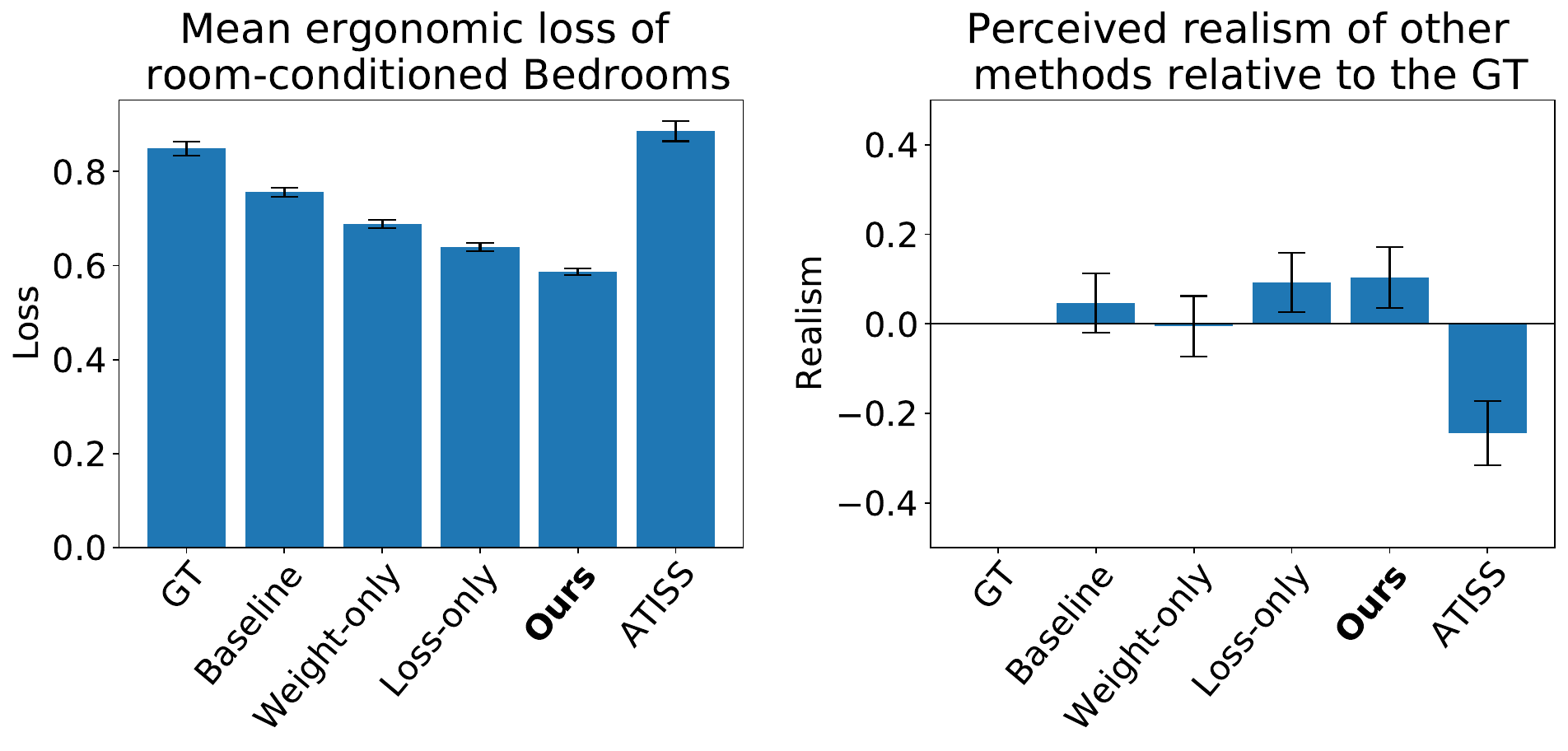}
    \caption{Room-conditioned \scene\ synthesis. We synthesize $20$ \scene\ variations for each floor plan in the Bedrooms validation set and evaluate the ergonomic \score. 
    The left chart shows the mean ergonomic loss of the synthesized \scenes, with the $80\%$ confidence interval of the mean shown in black. The realism of the synthesized \scenes\ is evaluated in a user study. %
    The right chart shows how the \scenes\ synthesized using each method are perceived compared to the ground truth, with a negative value meaning that the ground truth is seen as more realistic. Our proposed approach improves the ergonomic \score\ of the scenes, while also being perceived as more realistic than the ground truth.
    }
    \label{fig:user_study_ergo}
\end{figure}

We use our proposed model and its ablations introduced in the previous section for \scene\ synthesis and evaluate the results in terms of both realism and ergonomic loss. In order to evaluate the realism of our generated results, we perform a perceptual study using Amazon Mechanical Turk in which we ask participants to compare pairs of Bedroom \scenes\ with the question of which \scene\ is more realistic on a $7$-point scale. We compare \scenes\ from $6$ sources in this study: the ground truth \scenes\ from the 3DFRONT dataset \cite{fu2021future,fu2021front}, \scenes\ generated with our proposed model and its ablations, and another state-of-the-art method ATISS \cite{paschalidou2021atiss}, which we train using the code provided on their website, modified to include windows and doors in the same manner as our model. In each \scene\ pair, a synthesized \scene\ is compared to a ground truth \scene. A total of $330$ users participated in the study. Each pair of \scenes\ was shown $3$ times to $10$ different users each for a total of $30$ comparisons per \scene\ pair. %

The left side of the Figure \ref{fig:user_study_ergo} shows the mean ergonomic \score\ of all \scenes\ created for the user study. As can be seen, our approach performs the best at generating \scenes\ with lower ergonomic \score, reducing the mean ergonomic \score\ by $30.8\%$ compared to the ground truth data. The ablations of our model also improve the ergonomic \score\ to a lesser extend, including the baseline model which we attribute to our sampling strategy making it less likely to generate arrangements that are learned from outliers in the training data. On the other hand, \scenes\ created with ATISS show the highest ergonomic \score\ because the \scenes\ are perceived as less realistic than even our baseline model.

This can be seen on the right side of Figure \ref{fig:user_study_ergo} which shows how the users perceive the realism of synthesized \scenes\ compared to those of the ground truth in a range of $[-1,1]$, with a negative value meaning that the ground truth is seen as more realistic. The responses show that ATISS is considered significantly less realistic than the ground truth. On the other hand, the \scenes\ generated by all our models are seen as at least equally realistic as the ground truth \scenes, with users even preferring \scenes\ created with our full model over the ground truth. This shows that our approach can not only improve the ergonomic quality in a purely quantitative sense, but also improve the perceived realism of the \scenes.

A qualitative comparison is shown in Figure~\ref{fig:res:comp:gt_at_v3}. While all of the methods produce plausible layouts, our approach generates, on average, layouts with fewer ergonomic issues like missing light sources or poor accessibility.
Layouts sampled unconditionally for multiple room categories are shown in Figure~\ref{fig:res:ours:other_cat}. In these examples, all layout elements including the rooms, doors and windows are generated by the network. %

\newcommand{\scenewidth}{0.15}
\begin{figure}[!t]
\centering
     \rotatebox{90}{\hspace{10pt}Ours}
     \includegraphics[width=\scenewidth\textwidth,trim={40  100 40 30},clip]{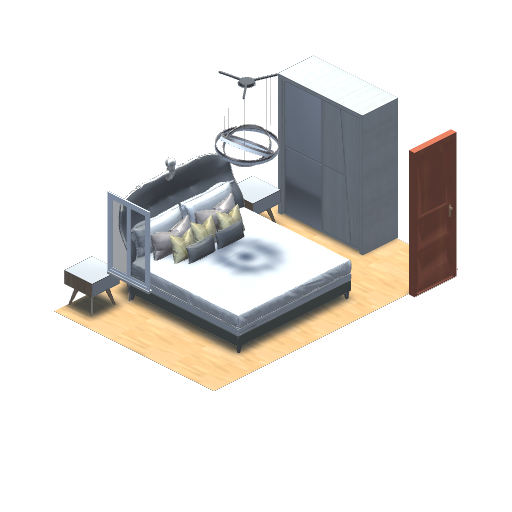} \hfill
     \includegraphics[width=\scenewidth\textwidth,trim={40  100 40 30},clip]{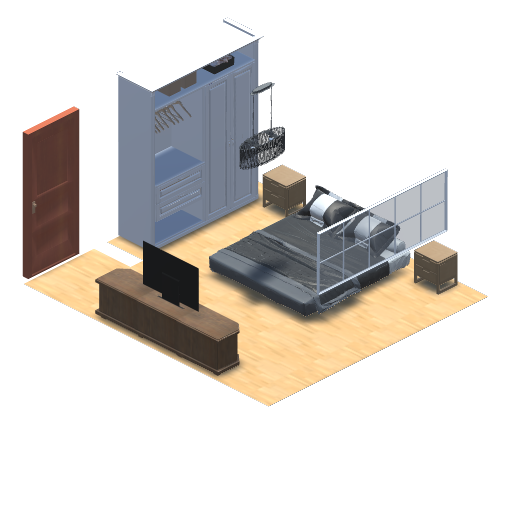}\hfill
     \includegraphics[width=0.15\textwidth,trim={40  100 40 30},clip]{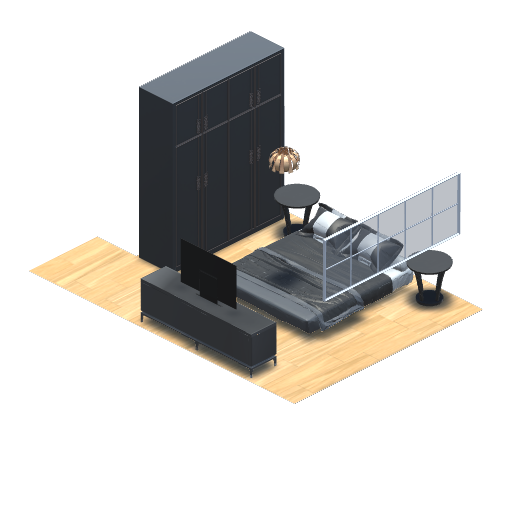}\hfill
    \\
     \rotatebox{90}{\hspace{10pt}Baseline}
     \includegraphics[width=\scenewidth\textwidth,trim={40  100 40 30},clip]{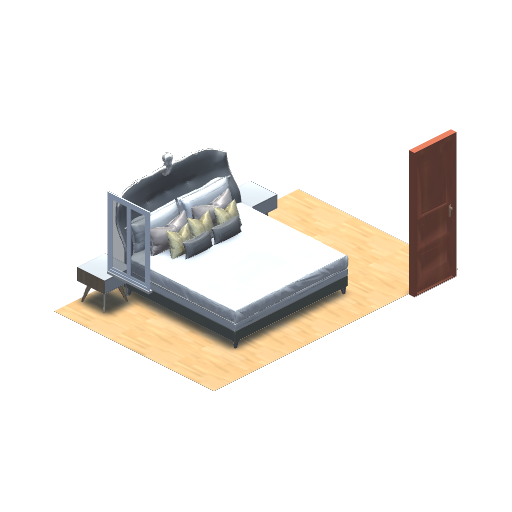} \hfill
     \includegraphics[width=\scenewidth\textwidth,trim={40  100 40 30},clip]{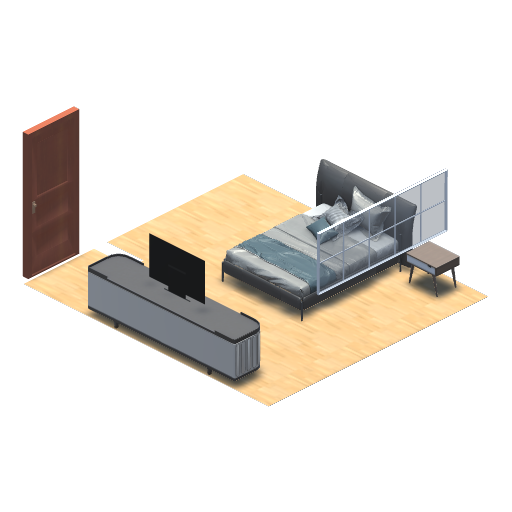}\hfill
     \includegraphics[width=0.15\textwidth,trim={40  100 40 30},clip]{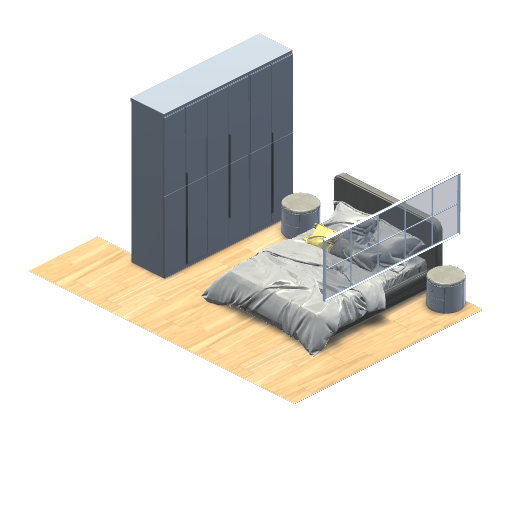}\hfill
    \\
    \rotatebox{90}{\hspace{15pt}ATISS}
     \includegraphics[width=\scenewidth\textwidth,trim={40  100 40 30},clip]{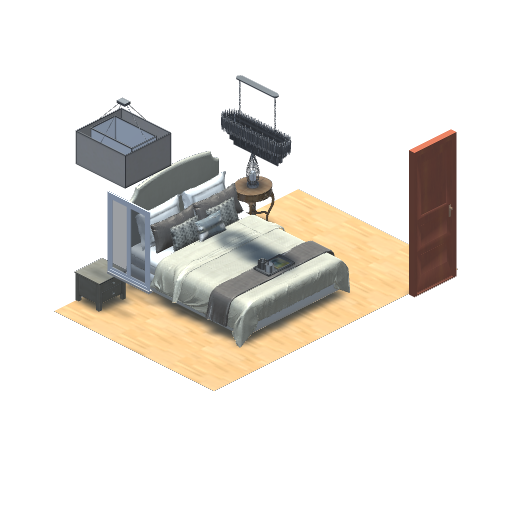} \hfill
     \includegraphics[width=\scenewidth\textwidth,trim={40  100 40 30},clip]{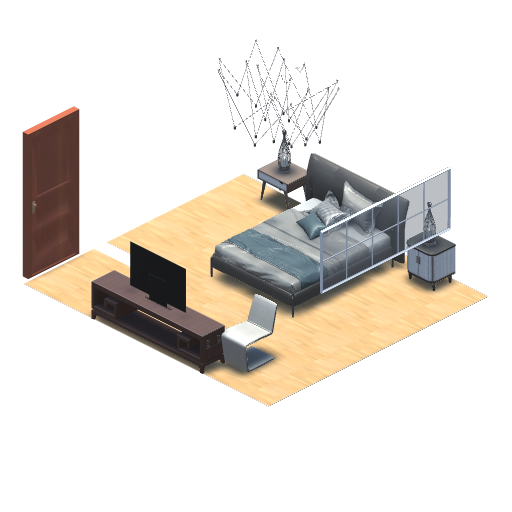}\hfill
     \includegraphics[width=0.15\textwidth,trim={40  100 40 30},clip]{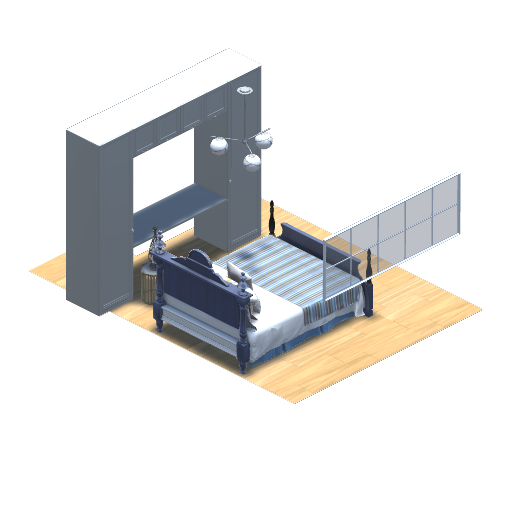}\hfill
    \\
     \rotatebox{90}{\hspace{10pt}Dataset}
     \includegraphics[width=\scenewidth\textwidth,trim={40  100 40 30},clip]{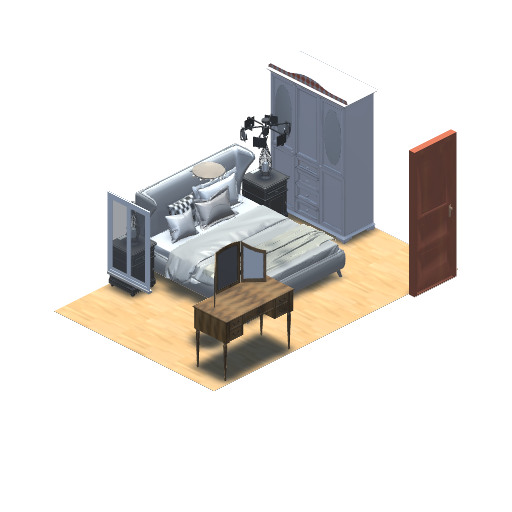} \hfill
     \includegraphics[width=\scenewidth\textwidth,trim={40  100 40 30},clip]{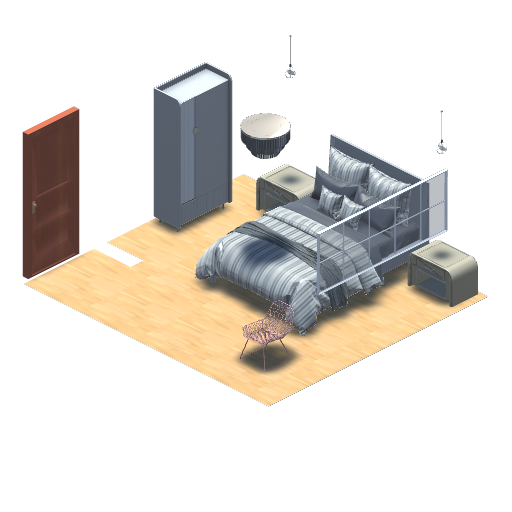}\hfill
     \includegraphics[width=0.15\textwidth,trim={40  100 40 30},clip]{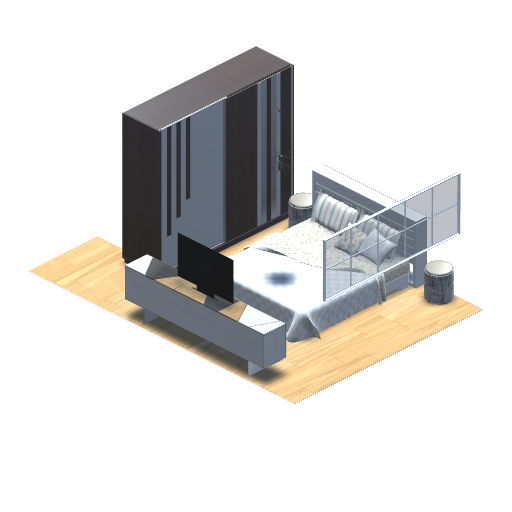}\hfill
    \caption{Conditional synthesis results
    as described in Section~\ref{sec:results}. Methods in a column receive the same room boundary, windows, and doors as input condition.
    Our approach produces on average \scene s with less ergonomic issues like missing light sources (e.g. Baseline columns 1, 2, 3)
    and poor accessibility (e.g. blocked path in ATISS column 3).
    }
    \label{fig:res:comp:gt_at_v3}
\vspace{-10pt}
\end{figure}

\begin{figure}[!t]%
\centering
     \rotatebox{90}{\hspace{10pt}Bedrooms}
     \includegraphics[width=\scenewidth\textwidth,trim={40  100 40 30}, clip]{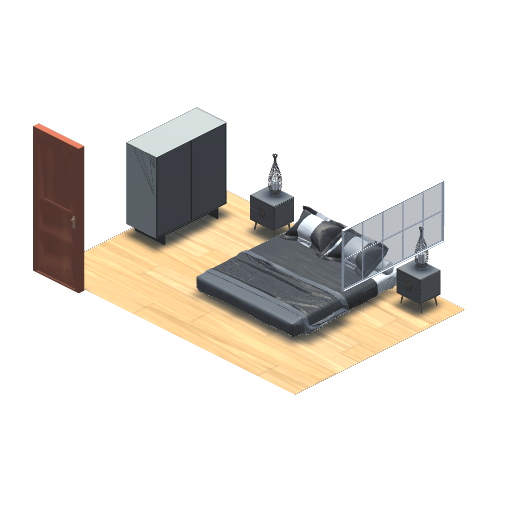}\hfill
     \includegraphics[width=\scenewidth\textwidth,trim={40  100 40 30}, clip]{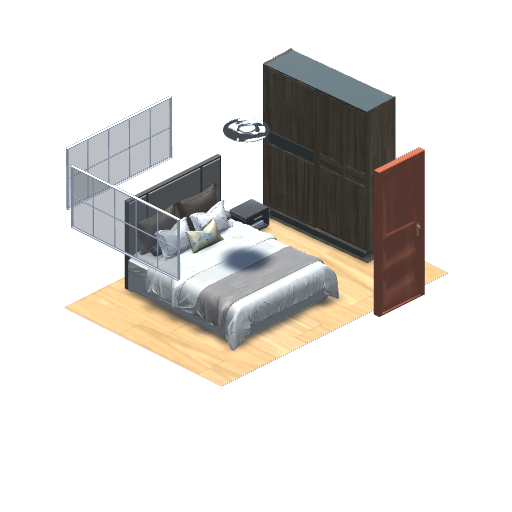}\hfill
     \includegraphics[width=\scenewidth\textwidth,trim={40  100 40 30}, clip]{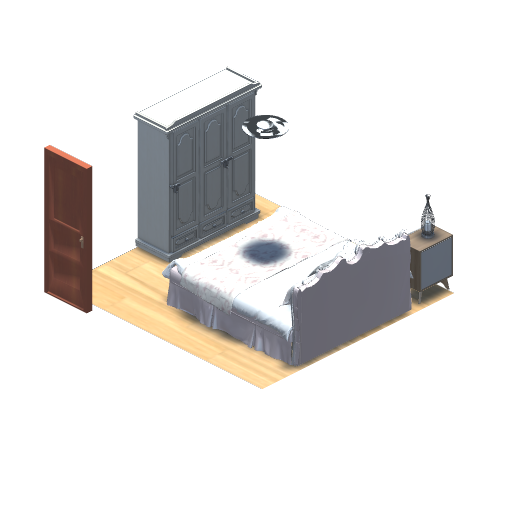}\hfill
     \\
     \rotatebox{90}{\hspace{5pt}Living Rooms}
     \includegraphics[width=\scenewidth\textwidth,trim={40  100 40 30},clip]{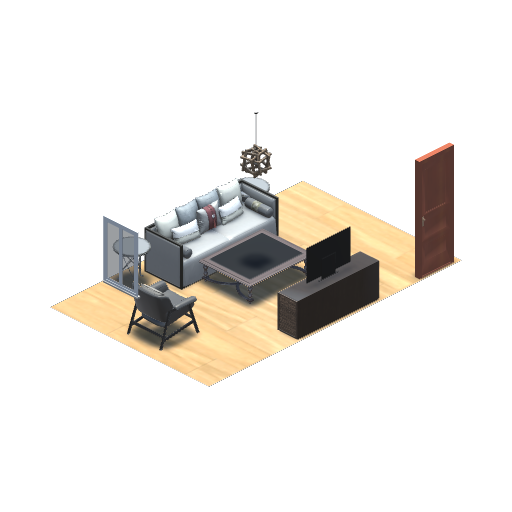}\hfill
     \includegraphics[width=\scenewidth\textwidth,trim={40  100 40 30},clip]{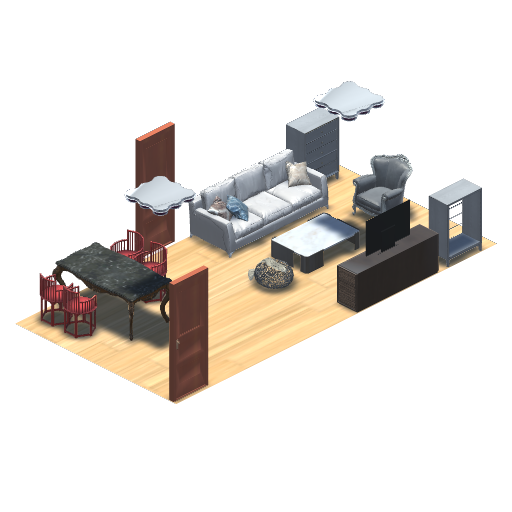}\hfill
     \includegraphics[width=\scenewidth\textwidth,trim={40  100 40 30},clip]{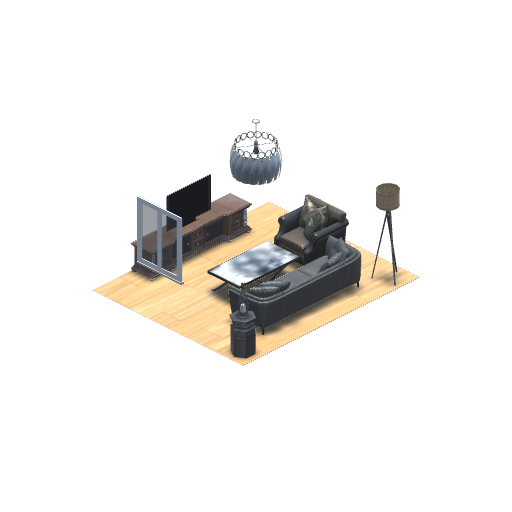}\hfill
     \\
     \rotatebox{90}{\hspace{5pt}Dining Rooms}
     \includegraphics[width=\scenewidth\textwidth,trim={40  100 40 30},clip]{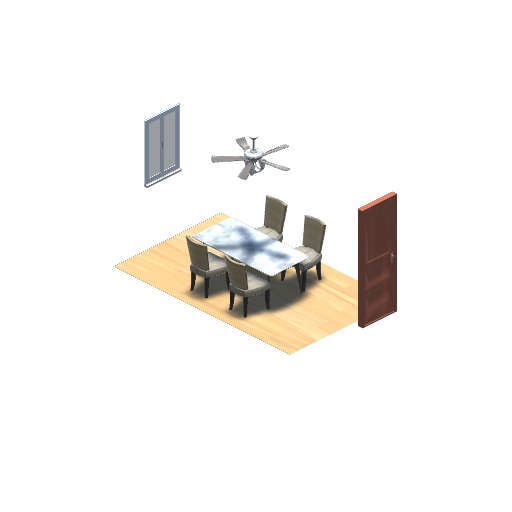}\hfill
     \includegraphics[width=\scenewidth\textwidth,trim={40  100 40 30},clip]{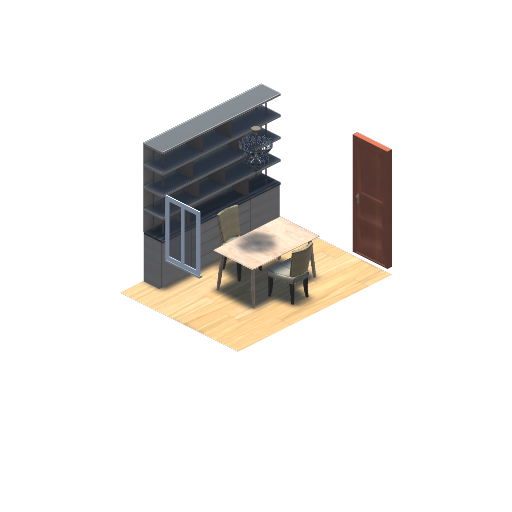}\hfill
     \includegraphics[width=\scenewidth\textwidth,trim={40  100 40 30},clip]{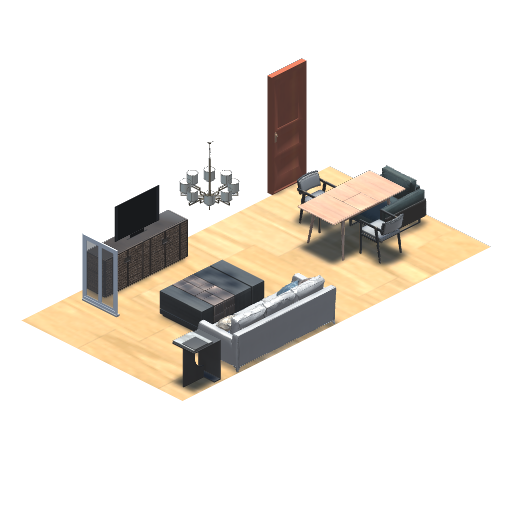}\hfill
     \rotatebox{90}{\hspace{20pt}Libraries}
     \includegraphics[width=\scenewidth\textwidth,trim={40  100 40 30}, clip]{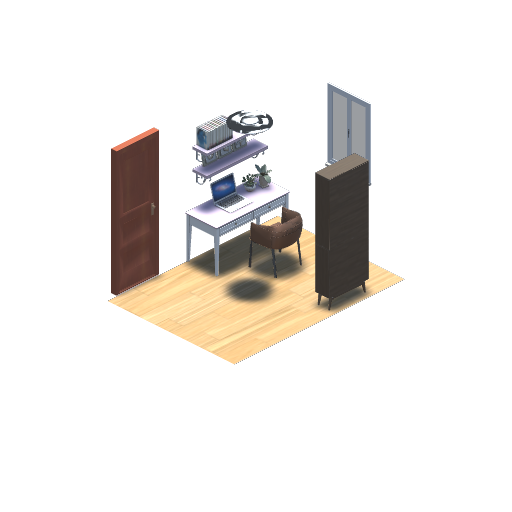}\hfill
     \includegraphics[width=\scenewidth\textwidth,trim={40  100 40 30}, clip]{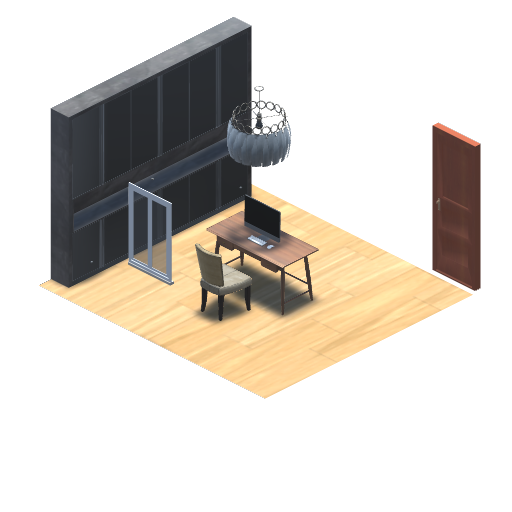}\hfill
     \includegraphics[width=\scenewidth\textwidth,trim={40  100 40 30}, clip]{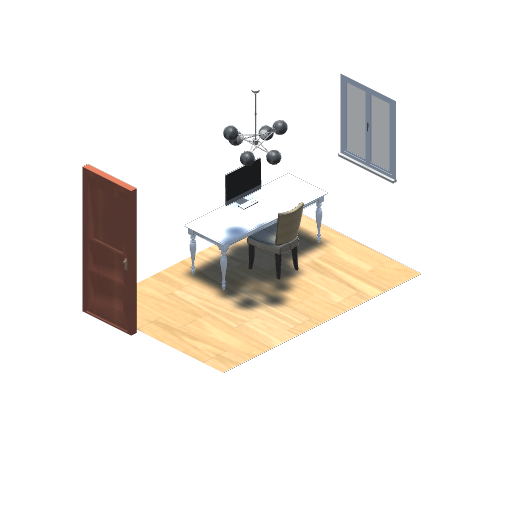}\hfill
    \caption{Generated \scene s for different room types. Since the attributes of the rooms were represented as part of the input sequences during training, all layout elements including rooms, doors, and windows can be generated by the network. Our method can generate furniture arrangements typical for each room type even with small training sets.
    }
    \label{fig:res:ours:other_cat}
\vspace{-10pt}
\end{figure}

\vspace{-2pt}
\section{Limitations and Conclusions}
\label{sec:discussion}

\subsection{Limitations} 
Our proposed approach has a number of limitations. Designing layouts is a complex high dimensional problem that includes modalities including selecting 3D furniture model that fit well together stylistically~\cite{weiss2020image,lun2015elements}; architectural elements such as room shapes walls and floor plans~\cite{wu2019data}; and various other aspects of lighting and illumination conditions~\cite{vitsas2020illumination}.
While important, such methods are orthogonal to our scope layout synthesis focused scope.

Furthermore, while our ergonomic loss functions are derived from ergonomics literature, they are only theoretical models and and have not been evaluated in a real-life setting. We think that the problem of translating the vast number of ergonomic rules and interior design guidelines into differentiable functions %
can be a promising topic of further research~\cite{schwartz2021human}.

While we have demonstrated that our approach of incorporating expert knowledge into the Transformer training process produces promising results, we think that this is only the first step in combining data-driven and rule-based learning using state-of-the-art deep-learning models such as Transformers. We believe that future research in this direction can assist with making data-driven learning approaches more applicable to domains where large amounts of high-quality data with desired properties are not readily available.

\subsection{Conclusions} 
We presented a novel method for the synthesis of indoor layouts, which combines data-driven learning and manually designed expert knowledge. To our knowledge, we are the first to propose such a solution to the problem. 
The main benefit of our approach is that it allows emphasizing features that might be underrepresented or not contained at all in the data. Simultaneously, we maintain the benefits of a data-driven approach which is important for layout generation which is high-dimensional and ill-defined. Manually crafting all design rules needed to synthesize comparable results would be very difficult and time consuming. Combining both expert knowledge and a distribution learned from data gives us the benefits from both worlds. 

As a technical contribution, we proposed a modern Transformer network that can be trained using a loss function composed of cross-entropy and additional knowledge. 
We have shown that weighting the two loss terms on a per-sample basis leads to results that fulfill the additional objective well and still maintain a high degree of realism.
Further, we introduced expert knowledge in the form of cost functions derived from ergonomics, whose goal is to improve layouts to be more usable and comfortable for humans. %

We described the details of our implementation (we will release our code on GitHub), and we evaluated the method thoroughly. %
We showed numerical quantitative results and performed a perceptual study %
where our model out-performs recent related work. We also used our system to synthesize a large set of realistically looking results. Our method is meant to help professionals and amateurs in the future to address the problem of interior layout design.

\begin{acks}
This research was funded by NJIT's faculty startup funds, 
gifts from Adobe, and by 
\grantsponsor{1}{Austrian Science Fund}{https://www.fwf.at}
\grantnum[https://www.fwf.at]{1}{FWF P 29981}.  
In addition, the authors would like to thank Stefan Pillwein for his help with paper production. 
\end{acks}

\balance
\bibliographystyle{ACM-Reference-Format}
\bibliography{refs}

\clearpage
\appendix

\renewcommand\thefigure{\thesection.\arabic{figure}}
\renewcommand{\thepage}{\thesection-\arabic{page}}

\section{Implementation Details}

\subsection{Ergonomic rules}\label{sec:scores:scores}

We consider the following ergonomic rules which are expressed as scalar cost functions in the range of $[0,1]$, where a lower value indicates a better score: (1) Reach, (2) Visibility, (3) Lighting, (4) Glare and (5) accessibility. In this section, we first describe the individual ergonomic cost functions for each rule, followed by the activities we use to evaluate the layouts.

\subsubsection{Reach}\label{sec:scores:reach}
While being seated, a person only has limited mobility and thus objects that need to be interacted with should be within a distance that is easy to reach without the need to stand up. We can broadly categorize the area around a seated person into $3$ zones. In the inner zone, objects can be reached without much effort, while objects in the outer zone are beyond reach. Objects in the middle zone can still be reached, but require more effort the further away they are. We model this reach cost $\loss{R}$ as a sigmoid function that measures how difficult it is to reach an object at position $q$ from position $p$:
\begin{equation}
	\loss{R} = \frac{1.0}{1.0 + \exp \left( -\beta_R  \left( \norm{q - p} - d_R \right) \right)} \,.
\end{equation}
The function is centered at $d_R$ with scaling parameter $\beta_R$. We use $d_R = 0.8$ and $\beta_R = 15$ to model the zones of easy and extended reach. These parameters roughly correspond to an easy reach up to $0.5m$ up to which the cost is close to $0$ and an extended reach up to $1.0m$, towards which the cost increases to $1.0$.%

\subsubsection{Visibility}\label{sec:scores:vis}
Visibility cost measures how visible an target object is from the viewpoint of the avatar given by position $p$ and viewing direction $u$. This measure is important for activities like watching TV or using the computer (cf. Table~\ref{tab:activities}), since seating furniture with sub-optimal positions or orientations may require the user to take on unhealthy postures.
To introduce this cost as smooth scalar function $E_v$ which can be minimized, we define the cost to increase with the angle between the two vectors $u$ and $v = \frac{q-p}{\norm{q-p}}$:
\begin{equation}
	\loss{V} = 1 - \left(\frac{ 1 + \langle u, v \rangle }{2} \right) \,.
\end{equation}

\subsubsection{Lighting}\label{sec:scores:light}
Lighting cost measures how well an object is illuminated by light sources in the room. Ideally, when looking at an object, the viewer and the light source should be positioned in the same half-space of the viewed object, as otherwise the object itself would partially obstruct the direct illumination and cause self-shadowing. A light source $b_i$ is thus well suited for illuminating the object at position $q$ when viewed from position $p$ as long as the position-to-object vector $v = \frac{q-p}{\norm{q-p}}$ and the vector $l_i = \frac{q-b_i}{\norm{q-b_i}}$ pointing from a light source at position $b_i$ to $q$ do not point in opposite directions:
\begin{equation*}
	e^L_i = \left(1 - \frac{ 1 + \langle v, l_i \rangle }{2} \right) \,.
\end{equation*}
Since multiple light sources can contribute to this cost, we compute their contribution by applying the $\softmin$ function to the vector $e^l = [e^l_i]_{i \in B}$ and using them as weights for computing the weighted sum:
\begin{equation}
	\loss{L} = \langle e^l, \softmin(\beta \cdot e^l) \rangle ,
\end{equation}
with $\beta$ being a temperature parameter that determines the hardness of the $\softmin$ function. We use $\beta = 10$. Since the computation of indirect illumination is prohibitively expensive, we only consider direct lighting. 

\subsubsection{Glare}\label{sec:scores:glare}
Glare cost $E_g$ measures the decrease in visual performance from strong brightness contrast caused by having bright light sources in the field of view. Given position-to-object vector $v = \frac{q-p}{\norm{q-p}}$ and glare vector $g_i = \frac{b_i-p}{\norm{b_i-p}}$ pointing from $p$ to the light source at $b_i$, the cost increases as the angle between the vectors decreases:
\begin{equation*}
	e^G_i = \left(\frac{ 1 + \langle v, g_i \rangle }{2} \right) \,.
\end{equation*}
Similar to the lighting cost we compute the weighted sum of multiple light sources using the $\softmax$ function for computing the weights:
\begin{equation}
	\loss{G} = \langle e^g, \softmax(\beta \cdot e^g) \rangle \,.
\end{equation}
For simplicity, we do not consider indirect glare, such as light sources that are reflected by a computer screen. Ceiling lights such as chandeliers are also excluded from this rule since light sources positioned above the field of view have a smaller impact on visual performance \cite{kroemer2017fitting}.

\subsubsection{Accessibility}\label{sec:scores:access}
The accessibility cost $\loss{A}$ measures how much space is available in front of a target object to allow easy interaction and walking through the room. For example, it is necessary to provide sufficient space between a bed and a wardrobe so that the wardrobe can be easily opened. We quantify this cost by defining an interaction region $I_j$ for each object $F_j$ that should not intersect with the bounding box $A_k$ of any another object $F_k$ in the layout. For most object categories, this region is located in front of the object itself, with a width equal to that of the object and an empirically chosen depth of $0.5m$. An exception is made for beds, since they are usually interacted with from the sides, so we define $2$ such regions on either side with a width equal to half the depth of the bed and a depth of $0.5m$. Given a \furnobj\ $F_j$ with interaction region $I_j$, we define the accessibility cost $E_a$ as
\begin{equation}
	\loss{A} = \sum_{k=0}^{N} \frac{\vert I_j \cap A_k \vert}{\vert I_j \vert} \,.
\end{equation}

\subsection{Activity Evaluation}

We evaluate the ergonomic \score\ of a \scene\ in the context of activities that are typically performed in rooms of a given category. Based on research on this topic \cite{puig2018virthome}, we select $4$ such activities which we label as \emph{Read book}, \emph{Watch TV}, \emph{Use computer} and \emph{Work at desk}. To evaluate an activity, it is necessary to compute the ergonomic costs relevant to that activity (cf. Table \ref{tab:activities}). We furthermore use a logarithmic function to re-scale the ergonomic cost functions to more strongly punish scenes with high costs, for example
\begin{equation}\label{eq:scaling}
	\bar{E}_{R} = -\ln (1.0 + \epsilon - E_{R}) ,
\end{equation}
with the scaling functions for the other rules defined analogously. We use $\epsilon = \exp(5)$, so that when $E_{R} = 1$, then $\bar{E}_{R} = 5$. We found this scaling function to be beneficial for minimizing the ergonomic loss during network training.

Since the accessibility cost $\loss{A}$ is relevant for every activity, we decide to compute this term once for the entire layout instead of computing it separately for every activity for performance reasons. We thus define the accessibility cost for the entire layout as 
\begin{equation}
    E_{access} = \langle e^{access}, \softmax(\beta \cdot e^{access}) \rangle ,
\end{equation}
with $e^{access} = [\bar{E}_A \left(I_j \right)]_{j=1,\ldots,N}$ being the vector containing the accessibility cost of every object and using $\beta = 10$. The $\softmax$ function is used to normalize the total cost of the layout such that a single badly-placed object increases the cost by roughly the same amount regardless of the number of objects in the layout.

\begin{table}[b]
  \centering
  \small
  \caption{Associations of rules to activities that can be performed in an environment. Not all activities require all rules to be fulfilled.  }
    \begin{tabular}{ p{6em}| c | c | c | c | c }
             & Reach & Visibility & Lighting & Glare & Accessibility\\
    \hline
    Read book &      &      & yes     & yes & yes \\
    \hline
    Watch TV &      & yes     &      & yes & yes \\
    \hline
    Use computer & yes     & yes     &      & yes & yes \\
    \hline
    Work at desk & yes     & yes     & yes     & & yes  \\
    \end{tabular}%
  \label{tab:activities}%
\end{table}%

For the activity \emph{Read book}, proper illumination conditions are the most important factor, so we need to apply the rules for lighting and glare. Given the position $p_j$ of seating furniture (like beds, chairs, or sofas), an associated object position $q_j$ (a book close to $p_j$) and light sources $B$ we define
\begin{equation*}
	e^{book}_j = \frac{\bar{E}_L \left(p_j,B,q_j \right) + \bar{E}_G \left(p_j,B,q_j \right)}{2} \,.
\end{equation*}
Since we do not require all possible positions to have a good score for every activity, we once again use the $\softmin$ function to compute a weighted sum of costs for the \scene. That way, if there is only one position that is suitable for an activity, it will be the only one with a large contribution to the \scene \,cost, while having multiple suitable positions will have them contribute equally. For a set of positions $p_j \in P$ we therefore have
\begin{equation}
	E_{book} = \langle e^{book}, \softmin(\beta \cdot e^{book}) \rangle ,
\end{equation}
with $e^{book} = [e^{book}_j]_{j \in P}$ and using $\beta = 10$.

The other activities are defined similarly. For \emph{Watch TV}, we require the TV to be visible from a piece of seating furniture and there should not be a light source in the field of view. We therefore compute the visibility and glare costs for positions $p_j$ with orientation $u_j$ (for chairs, beds, sofas) and TVs with position $q_k$:
\begin{equation*}
	e^{tv}_{j,k} = \frac{\bar{E}_V \left(p_j,u_j,q_k \right) + \bar{E}_G \left(p_j,B,q_k \right)}{2} \,.
\end{equation*}
Since there can be multiple TVs in a room in addition to multiple pieces of seating furniture, we need to compute the weighted sum of costs over every combination of $p_j$ and $q_k$, using $e^{tv} = [e^{tv}_{j,k}]_{j \in P,k \in Q}$:
\begin{equation}
	E_{tv} = \langle e^{tv}, \softmin(\beta \cdot e^{tv}) \rangle \,.
\end{equation}

The same rules are required for the activity \emph{Use computer}, in addition to the reach rule since the seating furniture and computer should be in close proximity. We do not evaluate the lighting rule because the direction from which the light illuminates the computer is not as important, since the computer screen is already illuminated. Using $q_k$ to denote the positions of computers we define 
\begin{equation*}
	e^{comp}_{j,k} = \frac{\bar{E}_V \left(p_j,u_j,q_k \right) + \bar{E}_G \left(p_j,B,q_k \right) + \bar{E}_R \left(p_j,q_k \right)}{3} \,.
\end{equation*}

Finally, for the activity \emph{Work at desk} we apply the rules visibility, lighting and reach. Since the viewing angle is mostly directed downward toward the desk during this activity, it is not necessary to consider direct glare caused by light sources in the room. Given chair positions $p_j$, table positions $q_k$ and light sources $B$ we compute
\begin{equation*}
	e^{work}_{j,k} = \frac{\bar{E}_V \left(p_j,u_j,q_k \right) + \bar{E}_L \left(p_j,B,q_k \right)  + \bar{E}_R \left(p_j,q_k \right)}{3} .
\end{equation*}

To obtain the overall \textbf{ergonomic loss} $E$ for a \scene\ we take the average of all activity costs that are possible in the \scene\ (e.g. if there is no computer in the scene, we do not evaluate the cost for \emph{Use computer}):
\begin{equation*}
	E = \frac{\sum_a \delta_{a} E_{a}}{\sum_a \delta_{a}} ,
\end{equation*}
with $a \in \braces{access,\ book,\ tv,\ comp,\ work}$ and $\delta_{a} = 1$ if the corresponding activity can be performed in the \scene\ and $\delta_{a} = 0$ otherwise.

\subsection{Dataset and Training Details}\label{sec:dataset}

\begin{figure*}[t]
    \centering
    \includegraphics[width=0.98\textwidth,trim={60  110 40 60},clip]{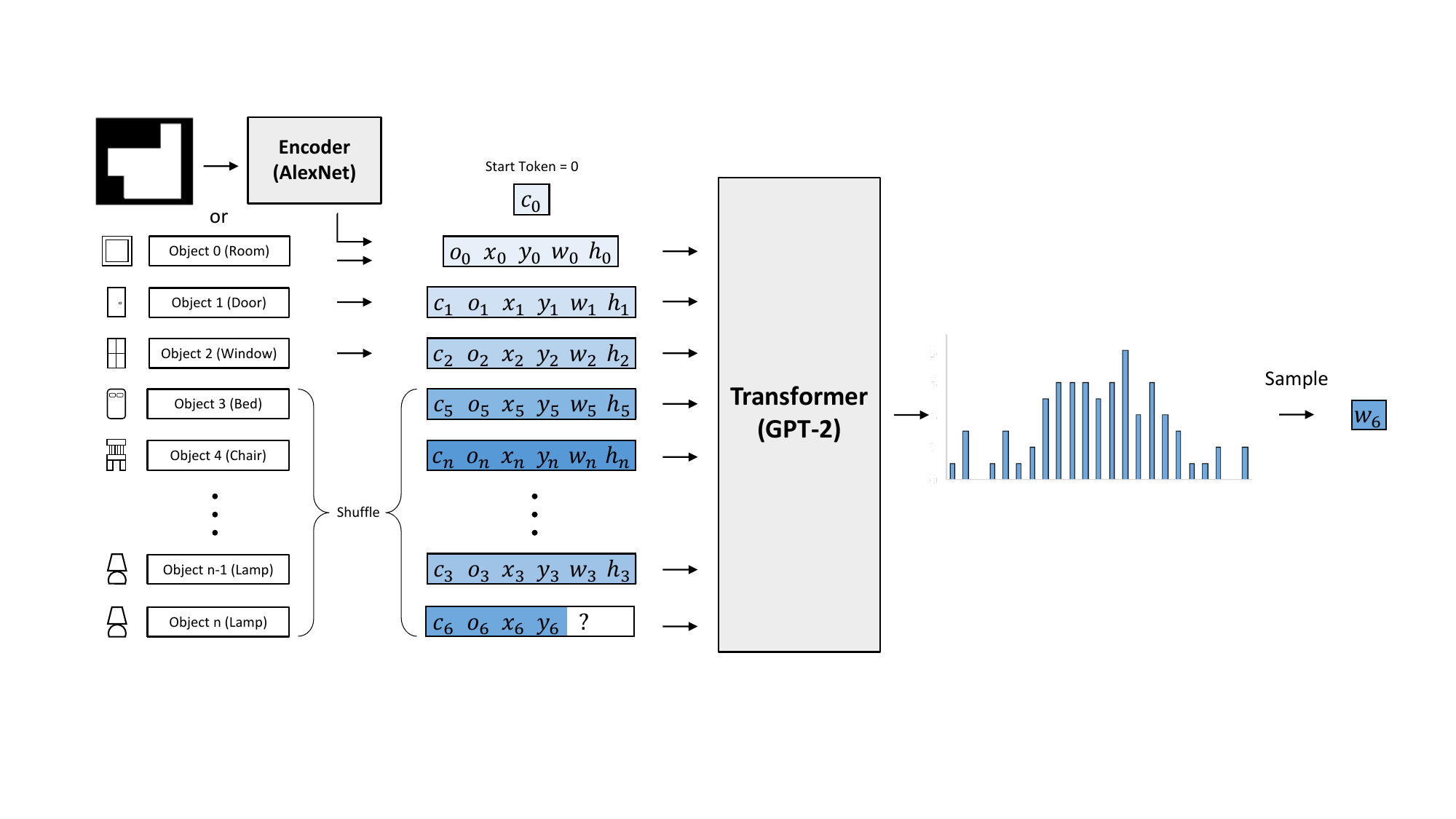}
    \caption{Overview of our model. A room layout consisting of individual furniture objects is mapped to a sequence of tokens which serves as the input to the transformer model. Given this sequence, the network predicts a categorical distribution for the next token from which we randomly sample the actual token value. During training, the order of objects other than the room, doors and windows is shuffled in the sequence. Furthermore, the attributes of the room can be either mapped to tokens directly (for rectangular rooms only), or by using an additional encoder network given a binary image of the floor plan as input.}
    \label{fig:process_overview}
\end{figure*}

\subsubsection{Dataset} \label{sub:results_dataset}

We train our models using the 3DFRONT dataset \cite{fu2021future,fu2021front} as training data. In a pre-processing step, we parse the data to extract rooms belonging to the categories Bedroom, Dining Room, Living Room and Library. For this purpose we use the filter criteria provided by ATISS \cite{paschalidou2021atiss}, consisting of a list of rooms for each category, as well as a split into training, testing and validation data. 
We use the rooms marked as \emphasis{train} for our training sets and combine those marked as \emphasis{test} and \emphasis{val} for our validation sets. Depending on the use case, we apply also some additional filtering. If we want to provide the attributes of the room shape as part of the input sequence, we can only use rectangular rooms and thus filter out rooms with more complex shapes. For the Bedrooms dataset, this results in $4041$ rooms for the training set and $324$ rooms for the validation set. For the direct comparison with ATISS, we provide the room shape using a binary map of the floor plan, allowing us to also use non-rectangular rooms, but we need to exclude rooms that have also been filtered by the pre-processing algorithm of ATISS. This leaves in $3526$ rooms for the training set and $289$ rooms for the validation set.

For most \furnobjs, their attributes such as the category and the transformation of the corresponding 3d model data can be directly extracted from the room data. Since separate 3d models for doors and windows are not provided with the dataset, we extract their positions and bounding box dimensions from the mesh data with corresponding labels. Since doors are only provided with each house and not attached to individual rooms, we include a door with the \furnobjs\ of a room if its distance to the closest wall of the room is lower than a chosen threshold and its orientation is aligned with that of the wall.

Additionally, we group some of the object categories in the dataset that are very similar to each other, while filtering out some others that occur only in very few rooms, for a total of $31$ categories that we use across all room types.

Since the dataset is typically lacking object categories that are necessary to properly evaluate the ergonomic \score\ of a \scene, we augment the dataset with additional objects in the following way. For each \scene, there is a $50\%$ chance to place a \furnobj\ of the indoor lamp category in the center of every stand and side-table object. In the same manner, a computer object is placed at the center of each desk object in a \scene\ with a probability of $50\%$. Finally, every TV stand object is augmented with a TV object.

\subsubsection{Training.} 
\begin{figure*}[t]
    \centering
    \includegraphics[width=0.98\textwidth]{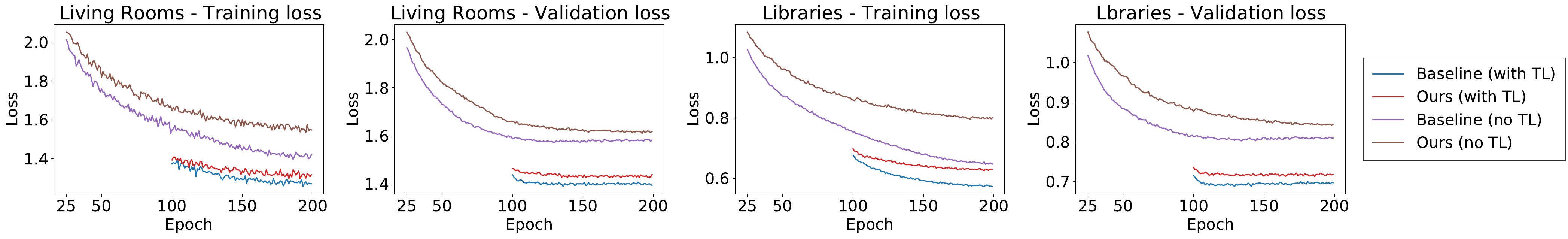}
    \caption{By pre-training the network on a general dataset containing samples from all room types and then fine-tuning the network for a specific room type, the validation loss can be decreased significantly, especially for small datasets.}
    \label{fig:transfer_learning}
\end{figure*}

As hyperparameters for our networks we use $12$ hidden layers, $8$ attention heads, embedding dimensionality of $256$, dropout probability of $0.1$ and a batch size of $64$.
Each network is trained for $150$ epochs, with the number of steps per epoch being equal to the number of training samples, such that each sample is used once per epoch. We use a learning rate of $0.0001$ with a linear rate of decay and a warm-up period of $10$ epochs. These parameters were determined empirically in preliminary experiments. For \scene\ synthesis, we always choose the learned network parameters of the epoch with the smallest validation loss during training.

When the shape of the room is provided as a binary map of the floor plan, we use an additional convolutional neural network to convert the binary map into a feature embedding. For this network we directly use the implementation of the AlexNet architecture~\cite{krizhevsky2012alexnet} provided in the code framework of ATISS~\cite{paschalidou2021atiss}. We also experimented with a ResNet-18 architecture~\cite{he2015resnet}, but found that AlexNet is more successful at discriminating between mirrored floor plans. The computed feature embedding then replaces the embedding that is otherwise computed from the orientation, position and dimension tokens of the room \furnobj.

Our networks are trained on Google Colab, using a machine with a NVIDIA Tesla P100 GPU. When only using the cross-entropy loss, training for one epoch takes $13$ seconds on average. Adding our ergonomic loss increases training times to $123$ seconds per epoch on average, since we cannot make use of parallelization for \scene\ evaluation as easily. There is room for further optimizations in this aspect.

\subsubsection{Inference.} During inference, we follow a similar approach to the strategy proposed by Sceneformer \cite{wang2020sceneformer}, using top-p nucleus sampling with $p=0.9$ for the object categories, as well as the attributes of the room, doors and windows. For the attributes of other object categories, we always pick the token with the highest probability.

The \scenes\ synthesized by the transformer network sometimes include intersecting objects which greatly disturb the perceived realism of a \scene. We therefore follow the approach of similar methods like Sceneformer and check for object intersections during inference. After the attributes of a \furnobj\ have been generated, we check if the object can be inserted into the scene without causing large intersections. If this is not the case, we re-sample the category and other attributes of the current object. If this re-sampling approach fails too often (we choose a limit of $20$ attempts experimentally), we discard the entire \scene\ and start anew. Certain pairs of object categories are excluded from this check, e.g. chairs can be put underneath a table and thus do not cause collisions.

In terms of computation time, the intersection-detection process is the bottleneck of the inference process. If we do check for intersection during inference, it takes $1653$ seconds for our models to synthesize $1000$ \scene sequences, for $1.653$ seconds per \scene \,on average. If we do not perform intersection-checks between objects, we can make use of parallelization to greatly reduce inference time. In such a setup, our networks can synthesize $1000$ \scene\ sequences in $27$ seconds for $0.027$ seconds per scene on average.

\subsubsection{Scene reconstruction.} Since our networks only generate the 2d bounding boxes of \furnobjs, we use an additional post-processing step to reconstruct a 3d scene from the generated \scene. For each \furnobj, we select the 3d model of the same category with the smallest difference in bounding box dimensions from the models in the 3DFRONT dataset~\cite{fu2021future,fu2021front}. For categories not included in the dataset, such as doors and windows, we handpick a few suitable models from online sources~\cite{turbosquid}. 

As a final step, the vertical position of each object is adjusted based on its category. The position of some categories like windows and chandeliers are set to a fixed height. We label some categories as supporting objects (like tables and stands) and others as supported objects (like indoor lamps and TVs). If there is an intersection between a supporting and supported object, the vertical position of the supported object is adjusted to be placed on top of the supporting object.

\subsection{Perceptual Study Setup}

\begin{figure}[b]
    \centering
    \includegraphics[width=0.48\textwidth]{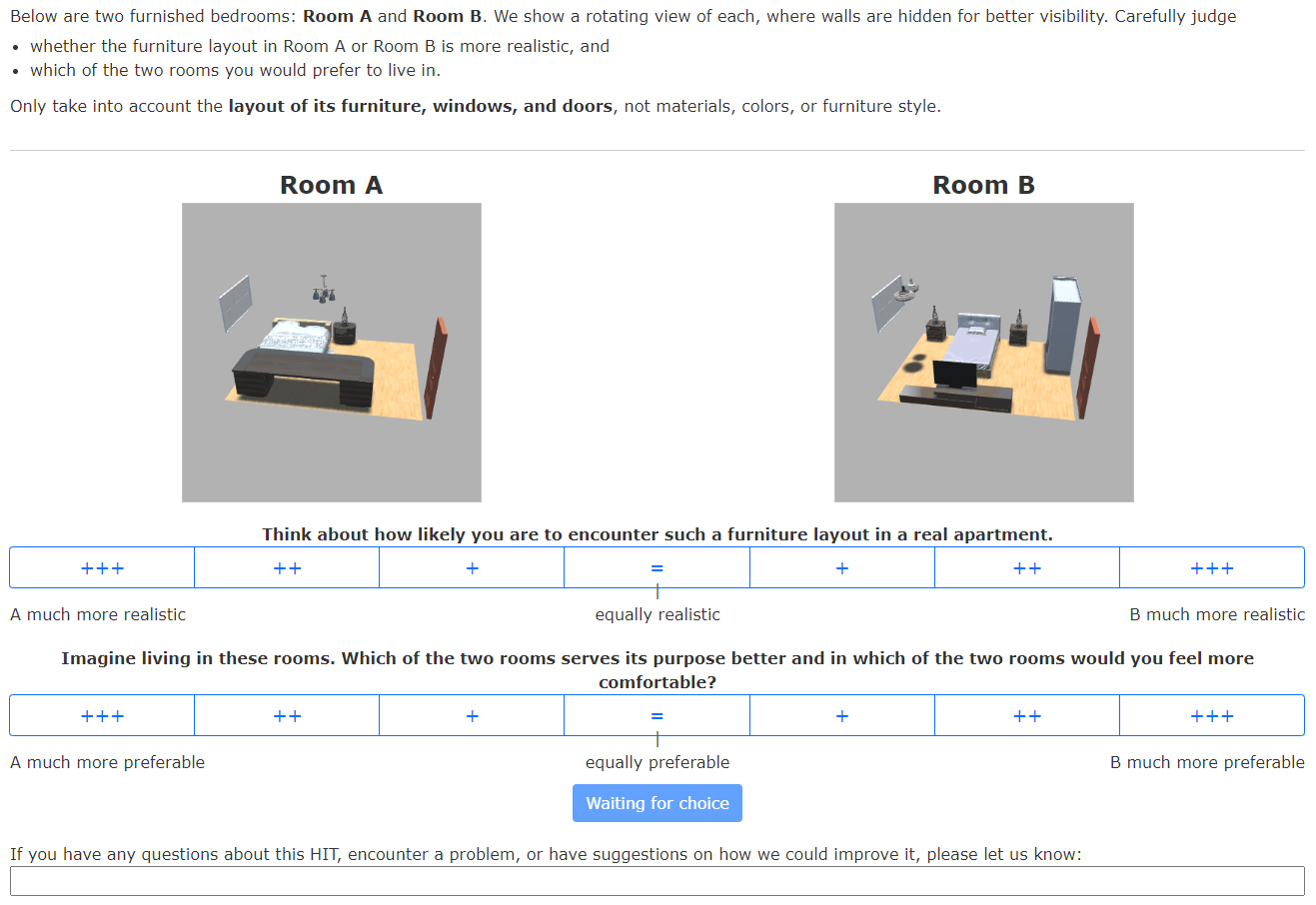}
    \caption{The interface of the user study. Participants were asked which of the $2$ displayed scenes is more realistic.}
    \label{fig:user_study_interface}
\end{figure}

In order to evaluate the realism of our generated results, we perform a perceptual study using Amazon Mechanical Turk in which we ask participants to compare pairs of Bedroom \scenes\ with the question of which \scene\ is more realistic.

To allow for a direct comparison, we use the binary map of the floor plan and the attributes of the doors and windows from the ground truth data for each \scene\ and only generate the rest of the \furnobjs\ using the selected methods. For each of the $289$ partial \scenes\ in the validation set consisting of binary map, doors, and windows, we generate $20$ furniture layout variations using both ATISS and our trained networks, resulting in a total of $5780$ sets of $6$ \scenes\ each, one for each method/variant. Since ATISS does not handle any intersections between \furnobjs\ and even some of the ground truth \scenes\ may contain such intersections, we discard the entire set if one of its \scenes\ contains an intersection between \furnobjs\ larger than a threshold, which we set as $20\%$ of the smaller bounding box area. For our networks, we perform intersection-checks during inference, only discarding a set if an intersection-free \scene\ cannot be generated after $20$ attempts. For comparison, $\sim 75\%$ of ATISS \scenes\ contain bounding box intersections, compared to $\sim 48\%$ of \scenes\ generated by our model with intersection check turned off. Additionally, $\sim 56\%$ of the ground truth \scenes\ also contain bounding box intersections. These numbers likely include some false positives where the bounding boxes intersect but the 3d meshes do not. We decided to be more conservative since intersections can significantly influence the perceived realism. Furthermore, since both ATISS and our model may try to generate additional windows or doors, we simple resample the category in such a case.  

For the user study, we randomly select $50$ sets from all sets of synthesized \scenes\ and ask users to compare the \scenes\ in terms of realism. In each comparison, the user is shown a pair of \scenes\ from the same set, each represented by an animated 3d rendering with the camera rotating around the scene. Users are asked which \scene\ is more realistic on a $7$-point scale (ranging from \emph{Scene A much more realistic} to \emph{Scene B much more realistic}, including a neutral \emph{Equally realistic} option). Figure~\ref{fig:user_study_interface} shows the screenshot of the UI. Each user sees the same scenes three times, and we use this redundancy to keep track of each user's consistency. We discard users that chose options more than two points apart for the same scene in more than $10\%$ of their comparisons. Additionally, we discard users that spent less than $10$ second on average per comparison. To evaluate the results, we compute a realism score for each method, that we obtain by assigning scores from $-1$ to $1$ to the $7$ possible user choices and averaging over all comparisons.

\subsection{Additional Room Types} \label{sec:res:transfer_learning}

Since some room types in the 3DFRONT dataset only contain few samples ($588$ living rooms, $554$ dining rooms and $424$ libraries for training set after our pre-processing), we make use of a transfer learning strategy. We first train a base model containing training samples of all room types for $100$ epochs using a learning rate of $\num{1e-4}$. This base model is then fine-tuned for each room type using a learning rate of $\num{2e-5}$ to prevent overfitting to the smaller datasets.

To evaluate the effectiveness of this approach, we train networks from scratch using only the training data from each individual room category and compare the cross-entropy loss to that of our networks which are first trained on a general set of training data before being fine-tuned for a room category. Figure~\ref{fig:transfer_learning} shows that the transfer learning strategy already yields a lower training and validation loss after the first epoch of fine-tuning. While the training loss for networks that are trained from scratch eventually approaches that of the pre-trained network, the validation loss remains higher throughout. As can be seen, for small training datasets, transfer learning proves to be a good strategy for improving the training process.

\section{Additional Qualitative Results}

Please see Figures \ref{fig:res:ours:other_cat_supp} and \ref{fig:res:comp:gt_at_v3_supp} for additional qualitative results supplementing those shown in the paper. Additionally, Figure~\ref{fig:res:comp:gt_at_v3} shows additional qualitative results for our model and its ablations.

\newcommand{\suppwidth}{0.14}

\begin{figure}[!t]
     \rotatebox{90}{\hspace{10pt}Bedrooms}
     \includegraphics[width=\suppwidth\textwidth,trim={40  100 40 30}, clip]{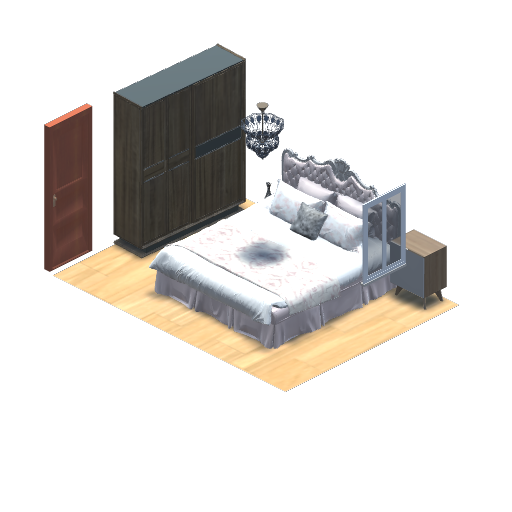}\hfill
     \includegraphics[width=\suppwidth\textwidth,trim={40  100 40 30}, clip]{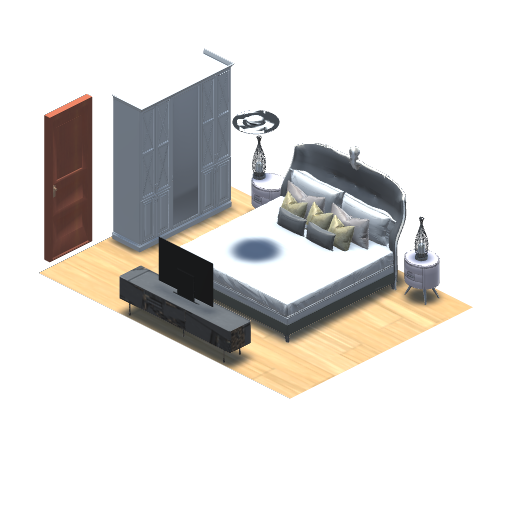}\hfill
     \includegraphics[width=\suppwidth\textwidth,trim={40  100 40 30}, clip]{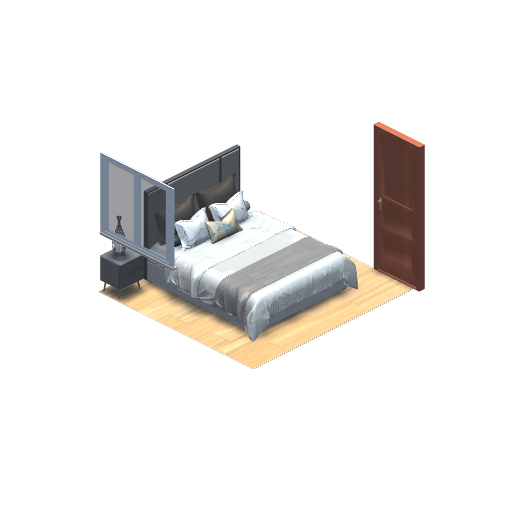}\hfill
     \\
     \rotatebox{90}{\hspace{5pt}Living Rooms}
     \includegraphics[width=\suppwidth\textwidth,trim={40  100 40 30},clip]{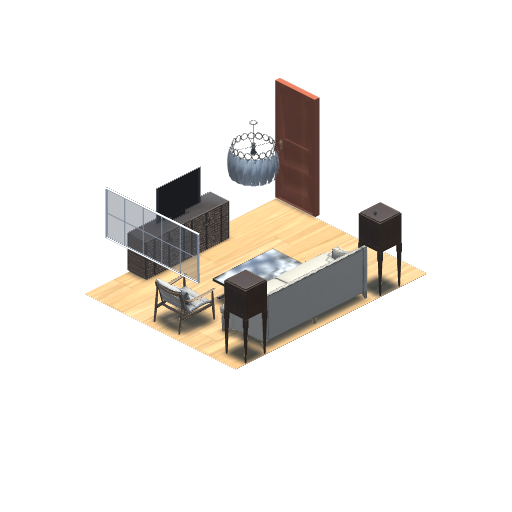}\hfill
     \includegraphics[width=\suppwidth\textwidth,trim={40  100 40 30},clip]{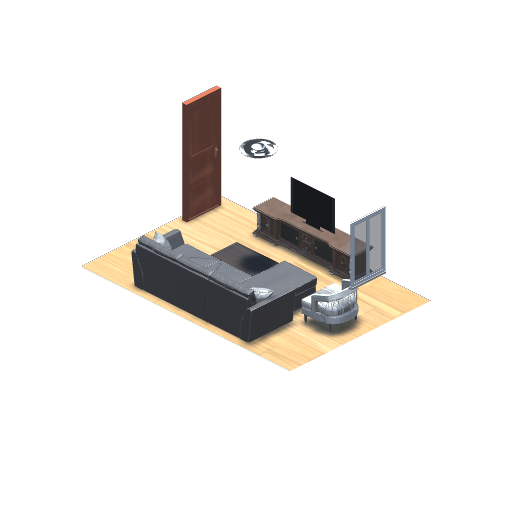}\hfill
     \includegraphics[width=\suppwidth\textwidth,trim={40  100 40 30},clip]{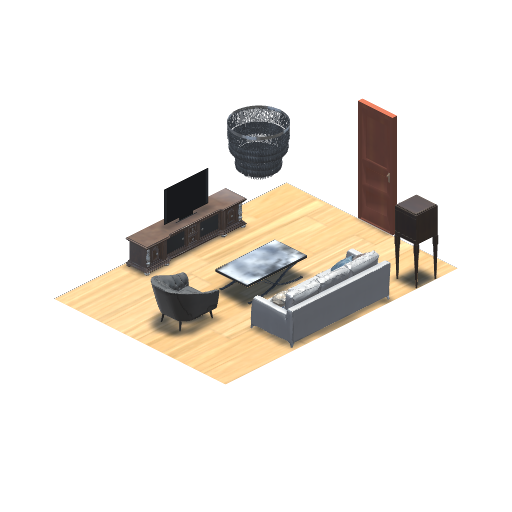}\hfill
     \\
     \rotatebox{90}{\hspace{0pt}Dining Rooms}
     \includegraphics[width=\suppwidth\textwidth,trim={40  100 40 30},clip]{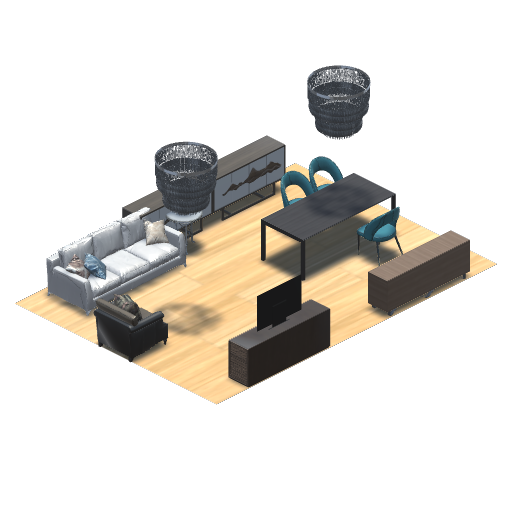}\hfill
     \includegraphics[width=\suppwidth\textwidth,trim={40  100 40 30},clip]{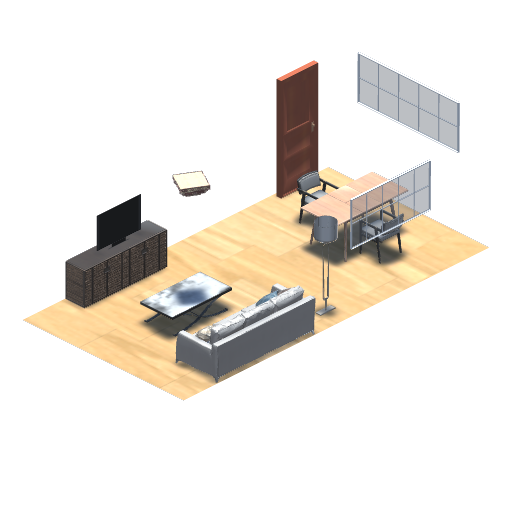}\hfill
     \includegraphics[width=\suppwidth\textwidth,trim={40  100 40 30},clip]{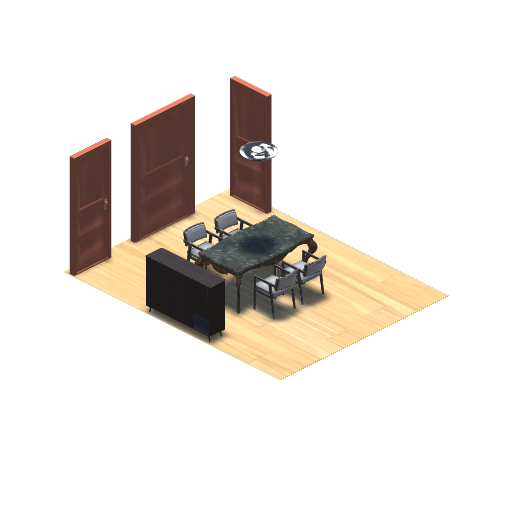}\hfill
     \\
     \rotatebox{90}{\hspace{12pt}Libraries}
     \includegraphics[width=\suppwidth\textwidth,trim={40  100 40 30}, clip]{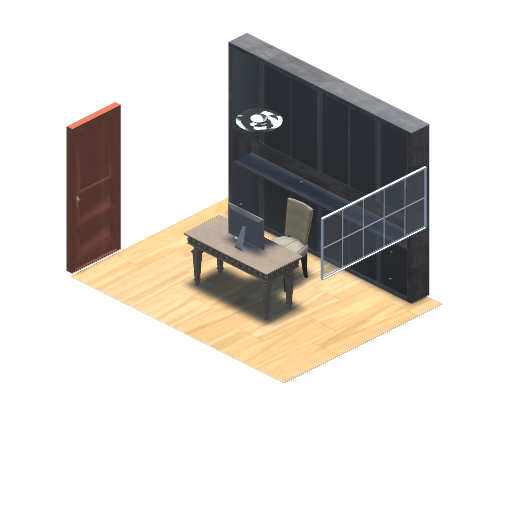}\hfill
     \includegraphics[width=\suppwidth\textwidth,trim={40  100 40 30}, clip]{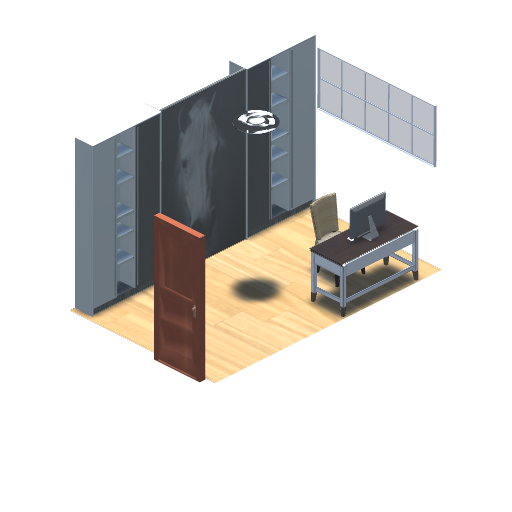}\hfill
     \includegraphics[width=\suppwidth\textwidth,trim={40  100 40 30}, clip]{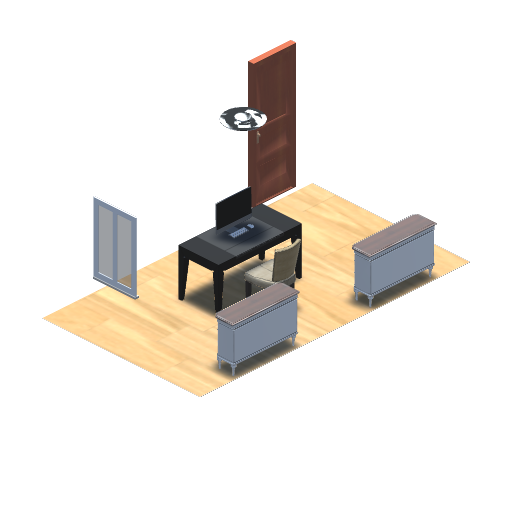}\hfill
    \caption{Additional generated \scene s for different room types. Since the attributes of the rooms were represented as part of the input sequences during training, all layout elements including rooms, doors, and windows can be generated by the network. Our method can generate furniture arrangements typical for each room type even with small training sets.
    }
    \label{fig:res:ours:other_cat_supp}
\end{figure}

\begin{figure}[!t]
     \rotatebox{90}{\hspace{15pt}Ours}
     \includegraphics[width=\suppwidth\textwidth,trim={40  100 40 30},clip]{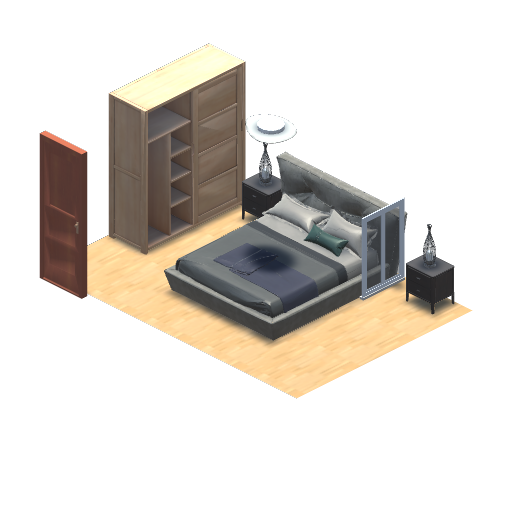}\hfill
     \includegraphics[width=\suppwidth\textwidth,trim={40  100 40 30},clip]{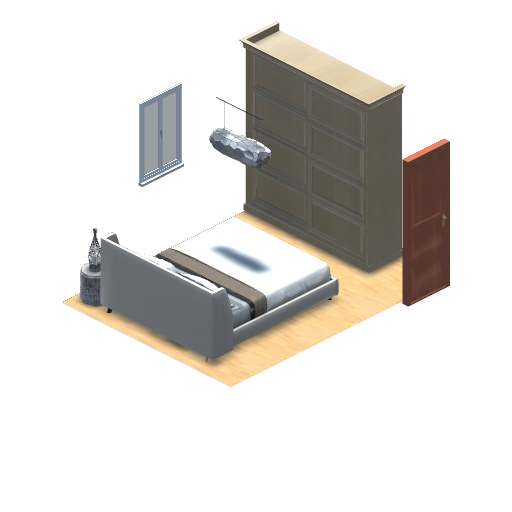}\hfill
     \includegraphics[width=\suppwidth\textwidth,trim={40  100 40 30},clip]{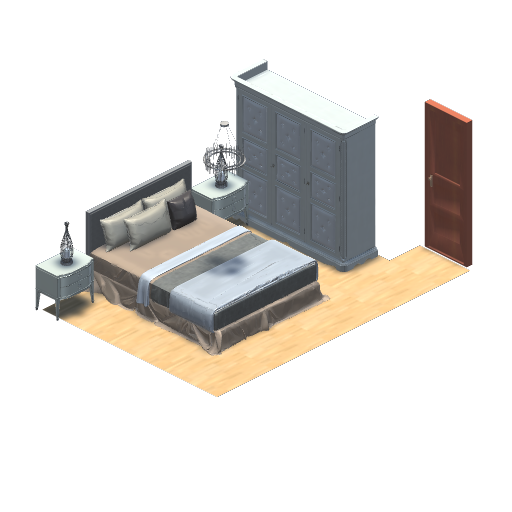}\hfill
    \\
     \rotatebox{90}{\hspace{12pt}Baseline}
     \includegraphics[width=\suppwidth\textwidth,trim={40  100 40 30},clip]{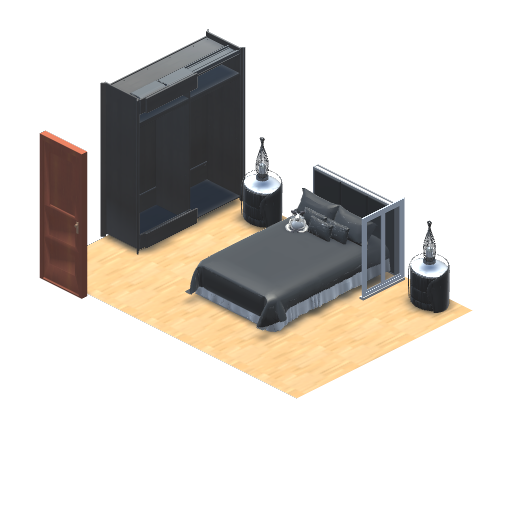}\hfill
     \includegraphics[width=\suppwidth\textwidth,trim={40  100 40 30},clip]{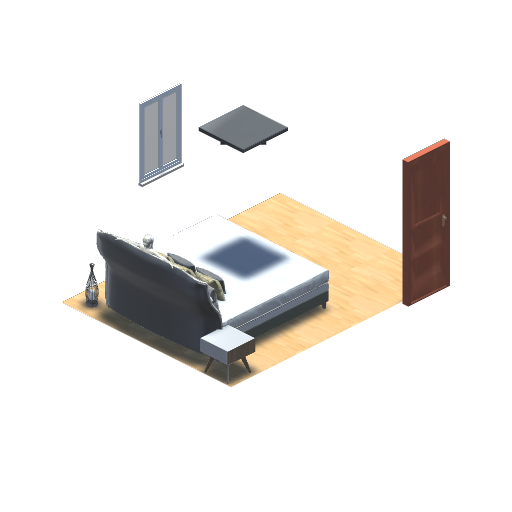}\hfill
     \includegraphics[width=\suppwidth\textwidth,trim={40  100 40 30},clip]{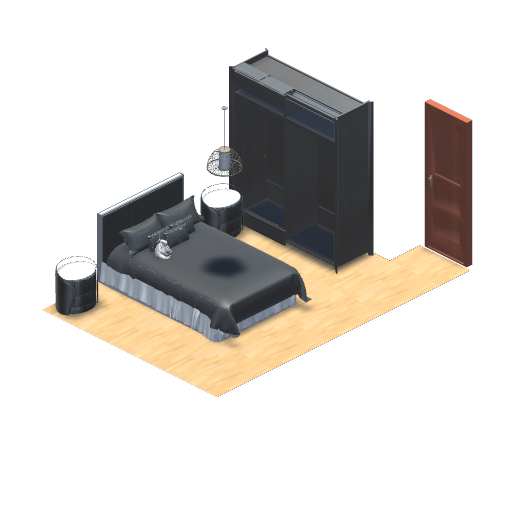}\hfill
    \\
    \rotatebox{90}{\hspace{15pt}ATISS}
     \includegraphics[width=\suppwidth\textwidth,trim={40  100 40 30},clip]{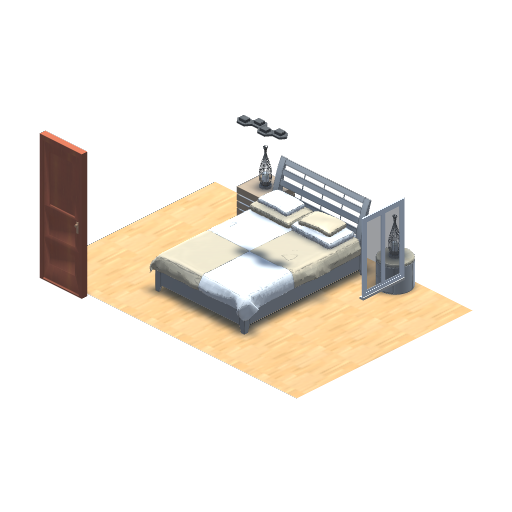}\hfill
     \includegraphics[width=\suppwidth\textwidth,trim={40  100 40 30},clip]{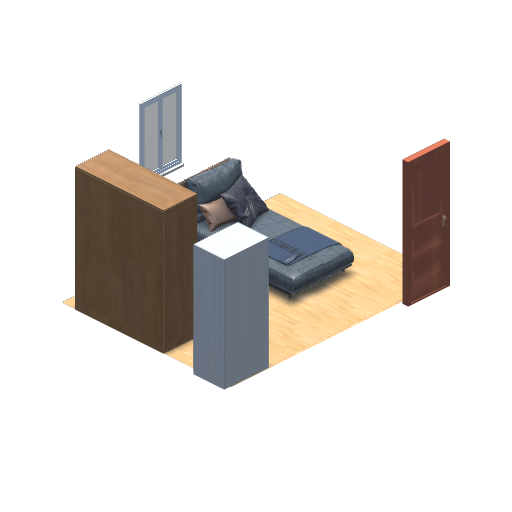}\hfill
     \includegraphics[width=\suppwidth\textwidth,trim={40  100 40 30},clip]{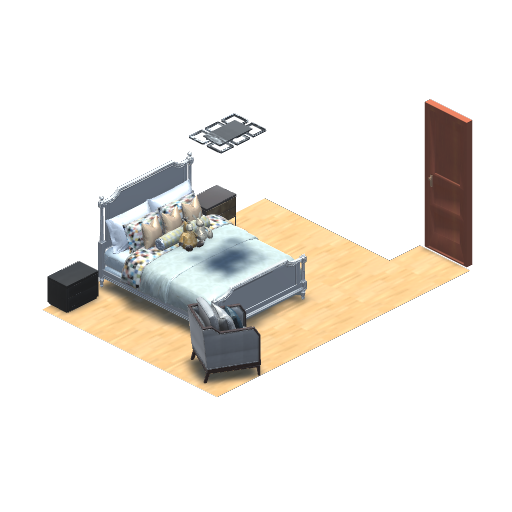}\hfill
    \\
     \rotatebox{90}{\hspace{12pt}Dataset}
     \includegraphics[width=\suppwidth\textwidth,trim={40  100 40 30},clip]{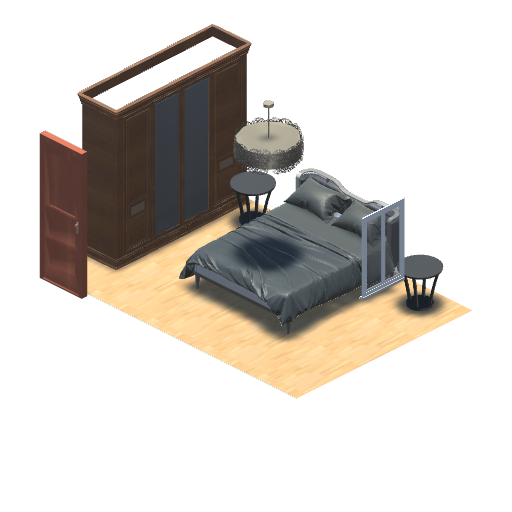}\hfill
     \includegraphics[width=\suppwidth\textwidth,trim={40  100 40 30},clip]{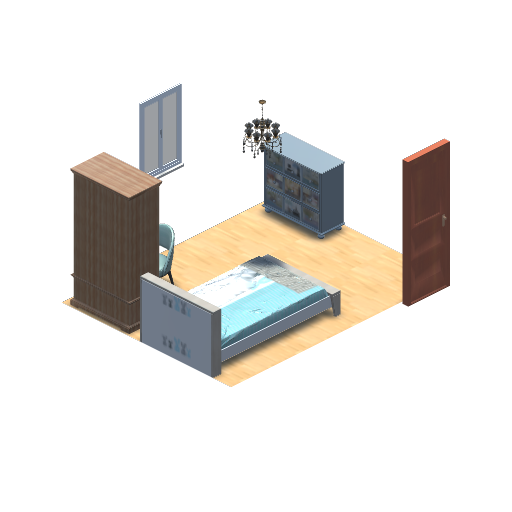}\hfill
     \includegraphics[width=\suppwidth\textwidth,trim={40  100 40 30},clip]{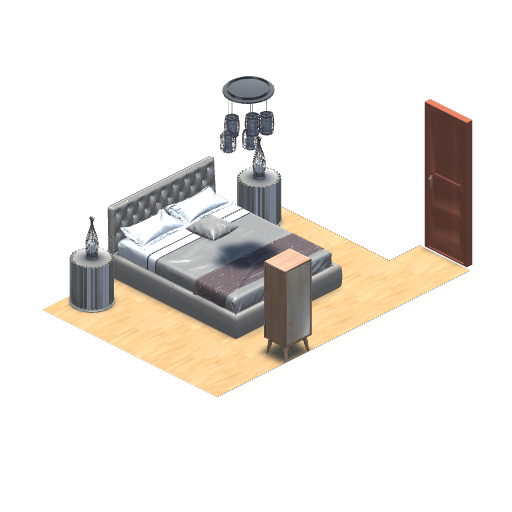}\hfill
    \caption{Additional qualitative comparisons of conditional synthesis results. Methods in a column receive the same room boundary, windows, and doors as input condition. Our approach produces on average \scene s with less ergonomic issues like missing light sources and poor accessibility.
    }
    \label{fig:res:comp:gt_at_v3_supp}
\end{figure}

\newcommand\figw{0.96}
\begin{figure*}[!t]
     \centering
     \begin{subfigure}[t]{\figw\textwidth}
         \centering
         \caption*{Results Ours (Full Model): Bedrooms}
         \includegraphics[width=0.15\textwidth,trim={0 80 0 80},clip]{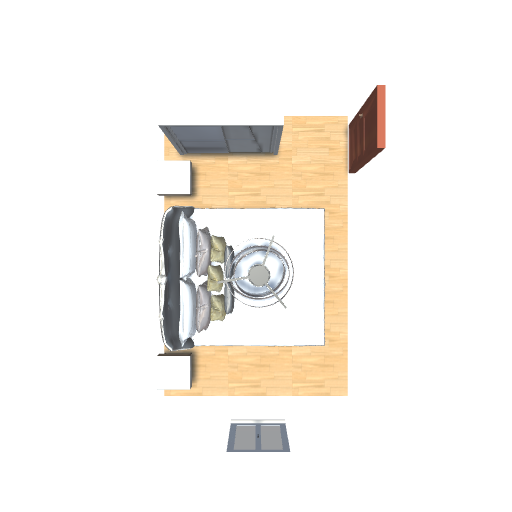}  \hfill
         \includegraphics[width=0.15\textwidth,trim={0 80 0 80},clip]{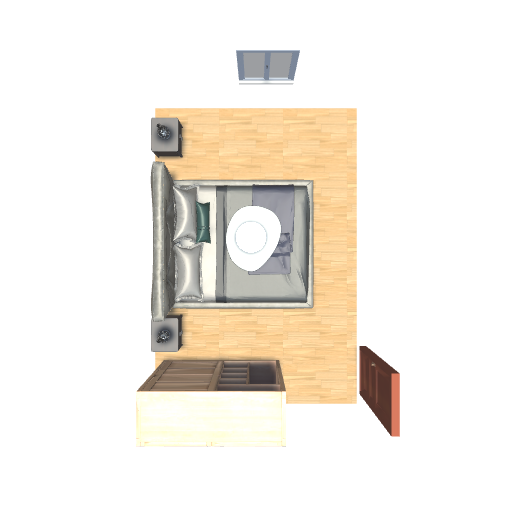} \hfill
         \includegraphics[width=0.15\textwidth,trim={0 80 0 80},clip]{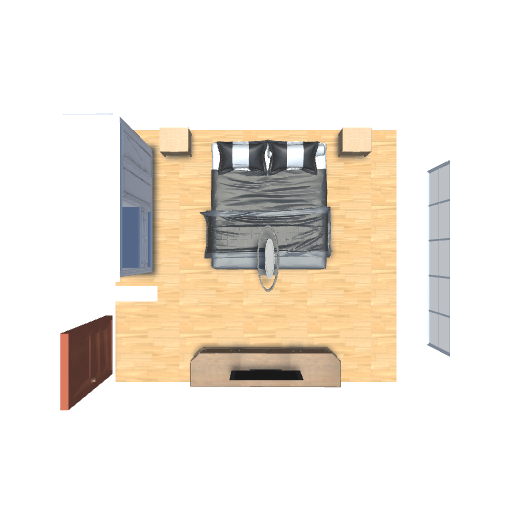} \hfill
         \includegraphics[width=0.15\textwidth,trim={0 80 0 80},clip]{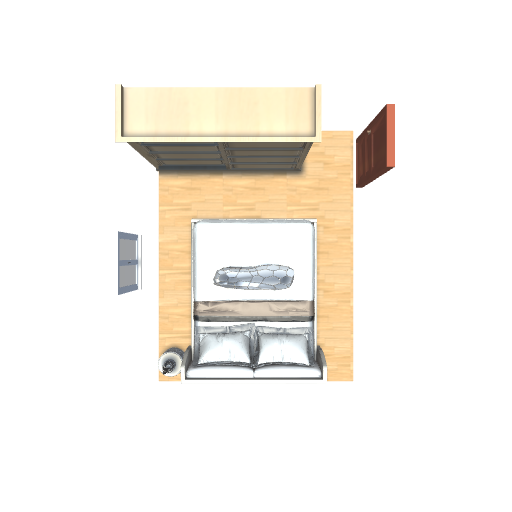} \hfill
         \includegraphics[width=0.15\textwidth,trim={0 80 0 80},clip]{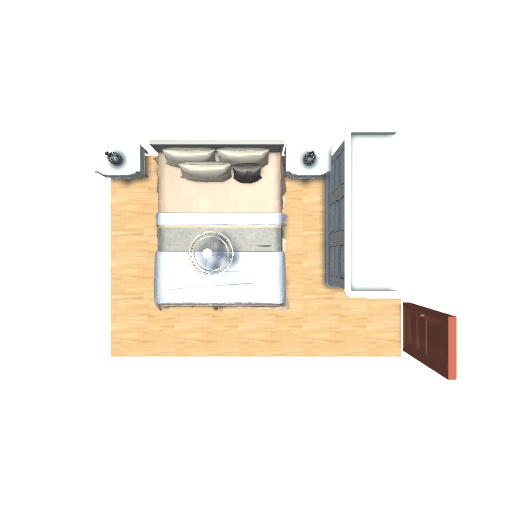} \hfill
         \includegraphics[width=0.15\textwidth,trim={0 80 0 80},clip]{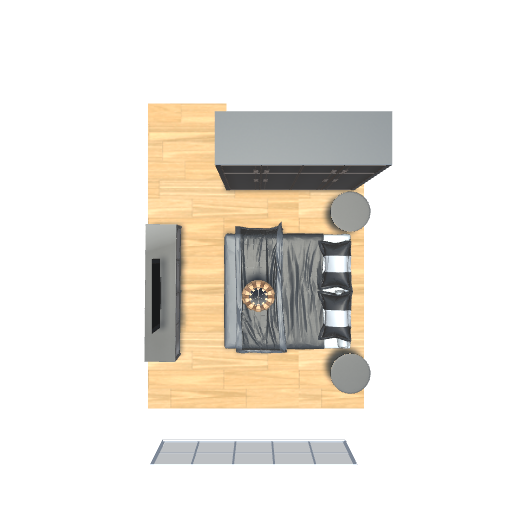} \hfill
         \includegraphics[width=0.15\textwidth,trim={40  100 40 30},clip]{images/renderings/main_512/Main_C3_Room2_4} \hfill
         \includegraphics[width=0.15\textwidth,trim={40  100 40 30},clip]{images/renderings/main_512/Main_C3_Room16_3}\hfill
         \includegraphics[width=0.15\textwidth,trim={40  100 40 30},clip]{images/renderings/main_512/Main_C3_Room21_4}\hfill
         \includegraphics[width=0.15\textwidth,trim={40  100 40 30},clip]{images/renderings/main_512/Main_C3_Room23_4}\hfill
         \includegraphics[width=0.15\textwidth,trim={40  100 40 30},clip]{images/renderings/main_512/Main_C3_Room26_1}\hfill
         \includegraphics[width=0.15\textwidth,trim={40  100 40 30},clip]{images/renderings/main_512/Main_C3_Room31_1}\hfill
         
         \label{fig:res_v3}
     \end{subfigure}
     \begin{subfigure}[t]{\figw\textwidth}
         \centering
         \caption*{Results Ours (Baseline): Bedrooms}
         \includegraphics[width=0.15\textwidth,trim={0 80 0 80},clip]{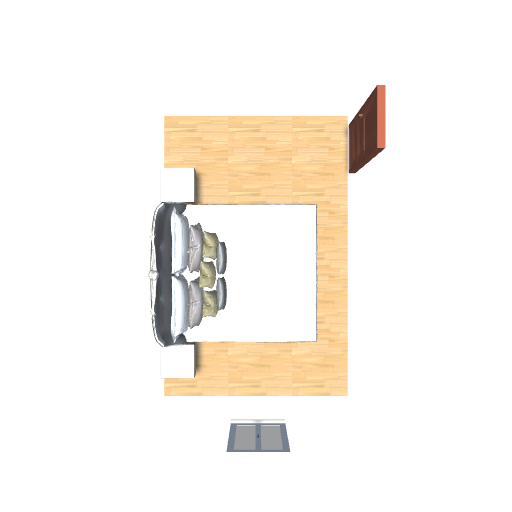} \hfill
         \includegraphics[width=0.15\textwidth,trim={0 80 0 80},clip]{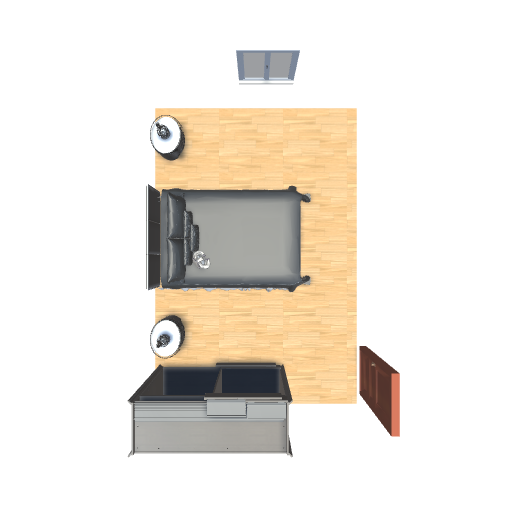}\hfill
         \includegraphics[width=0.15\textwidth,trim={0 80 0 80},clip]{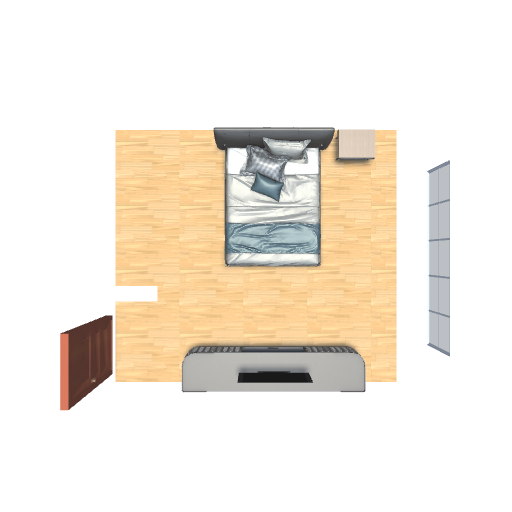}\hfill
         \includegraphics[width=0.15\textwidth,trim={0 80 0 80},clip]{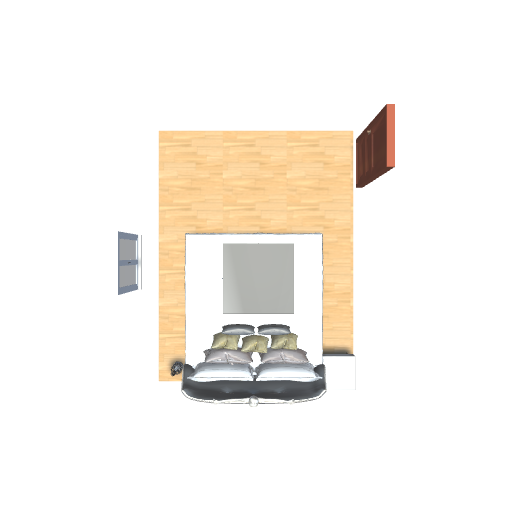}\hfill
         \includegraphics[width=0.15\textwidth,trim={0 80 0 80},clip]{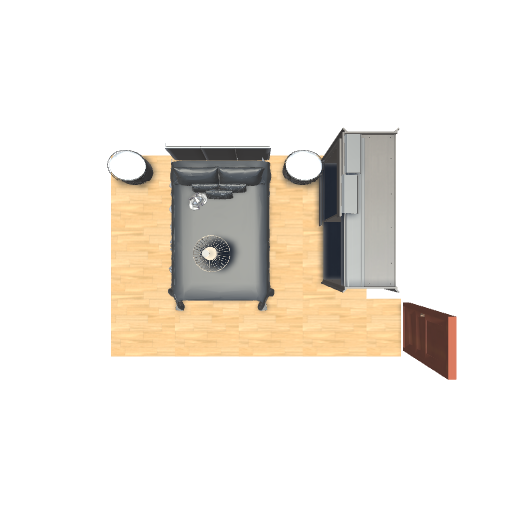}\hfill
         \includegraphics[width=0.15\textwidth,trim={0 80 0 80},clip]{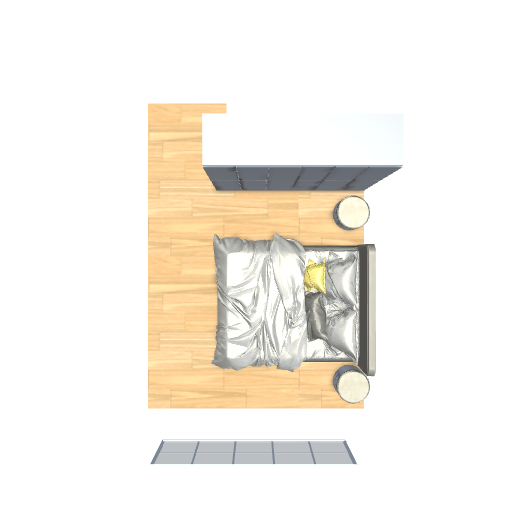}\hfill
         \includegraphics[width=0.15\textwidth,trim={40  100 40 30},clip]{images/renderings/main_512/Main_C0_Room2_4} \hfill
         \includegraphics[width=0.15\textwidth,trim={40  100 40 30},clip]{images/renderings/main_512/Main_C0_Room16_3}\hfill
         \includegraphics[width=0.15\textwidth,trim={40  100 40 30},clip]{images/renderings/main_512/Main_C0_Room21_4}\hfill
         \includegraphics[width=0.15\textwidth,trim={40  100 40 30},clip]{images/renderings/main_512/Main_C0_Room23_4}\hfill
         \includegraphics[width=0.15\textwidth,trim={40  100 40 30},clip]{images/renderings/main_512/Main_C0_Room26_1}\hfill
         \includegraphics[width=0.15\textwidth,trim={40  100 40 30},clip]{images/renderings/main_512/Main_C0_Room31_1}\hfill
         \label{fig:res_v0}
     \end{subfigure}
     \begin{subfigure}[t]{\figw\textwidth}
         \centering
         \caption*{Results Ours (Weight-Only): Bedrooms}
         \includegraphics[width=0.15\textwidth,trim={0 80 0 80},clip]{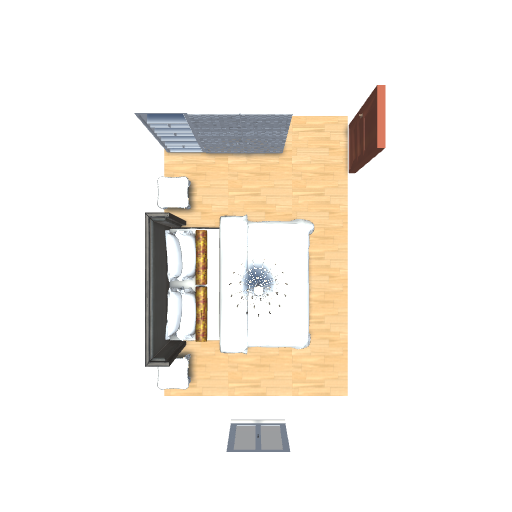} \hfill
         \includegraphics[width=0.15\textwidth,trim={0 80 0 80},clip]{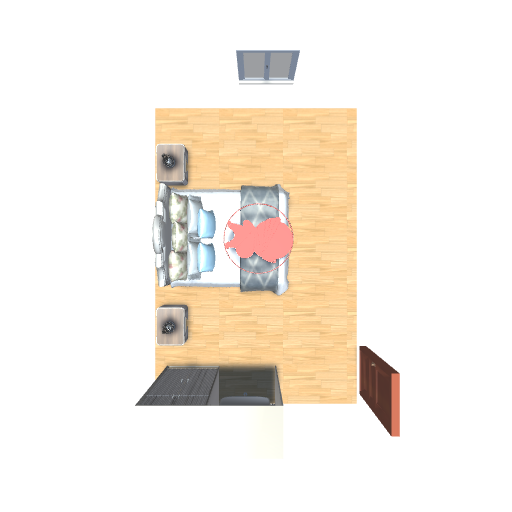}\hfill
         \includegraphics[width=0.15\textwidth,trim={0 80 0 80},clip]{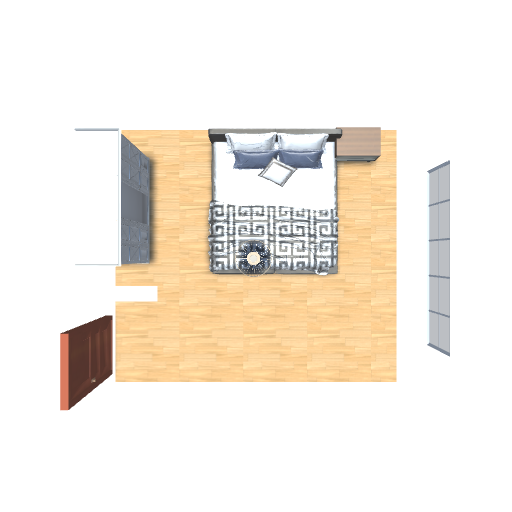}\hfill
         \includegraphics[width=0.15\textwidth,trim={0 80 0 80},clip]{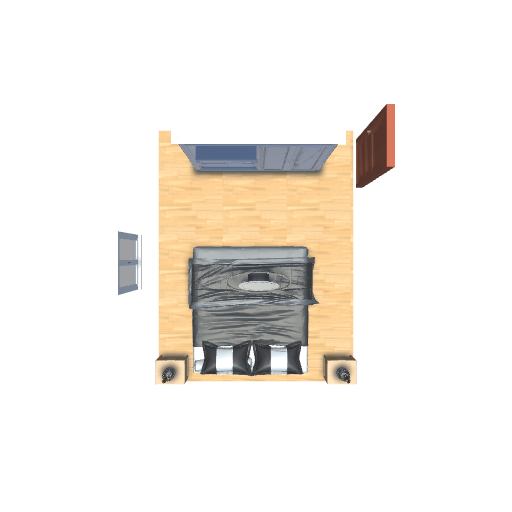}\hfill
         \includegraphics[width=0.15\textwidth,trim={0 80 0 80},clip]{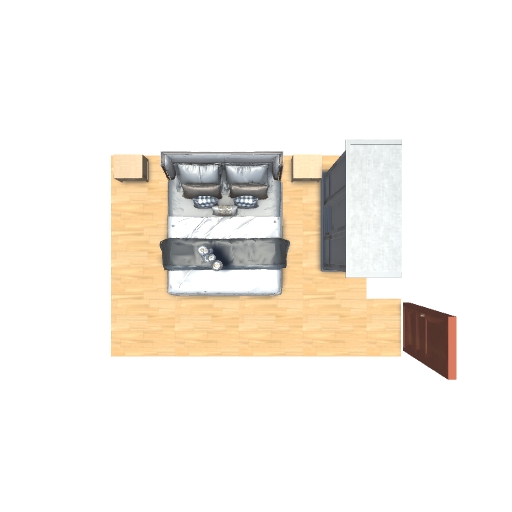}\hfill
         \includegraphics[width=0.15\textwidth,trim={0 80 0 80},clip]{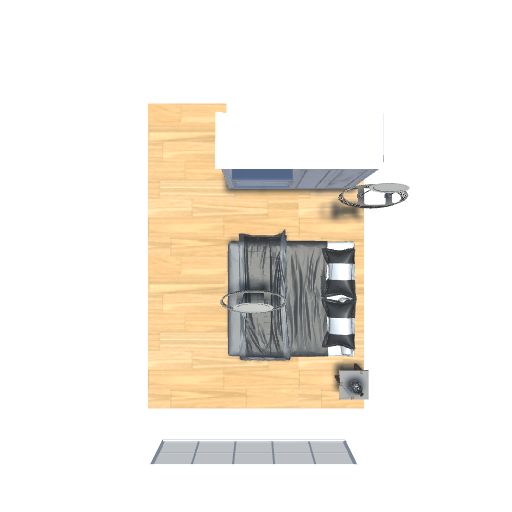}\hfill
         \includegraphics[width=0.15\textwidth,trim={40  100 40 30},clip]{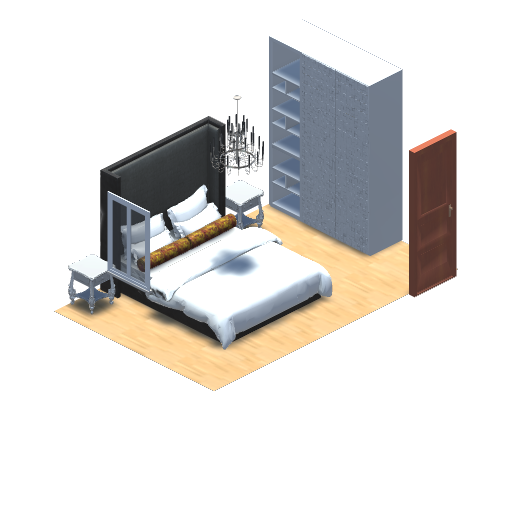} \hfill
         \includegraphics[width=0.15\textwidth,trim={40  100 40 30},clip]{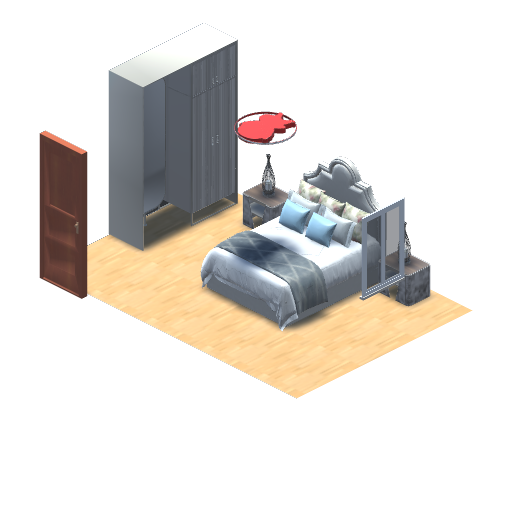}\hfill
         \includegraphics[width=0.15\textwidth,trim={40  100 40 30},clip]{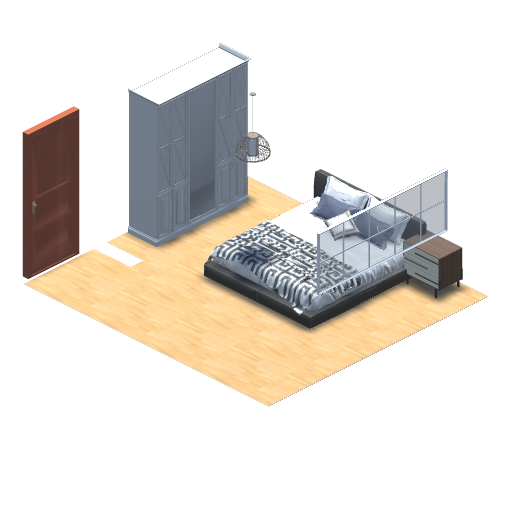}\hfill
         \includegraphics[width=0.15\textwidth,trim={40  100 40 30},clip]{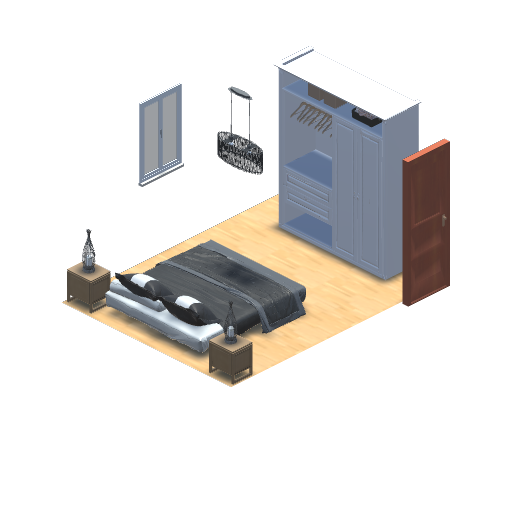}\hfill
         \includegraphics[width=0.15\textwidth,trim={40  100 40 30},clip]{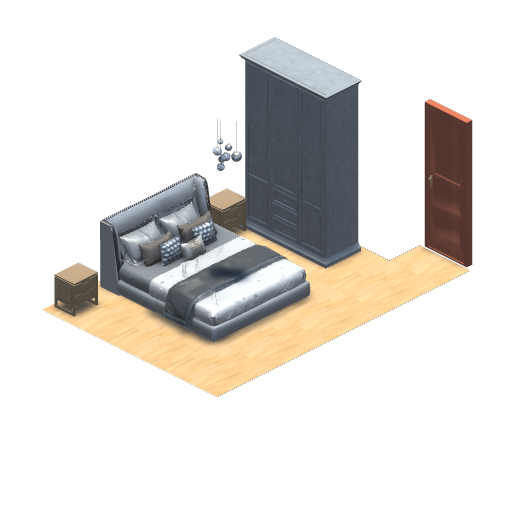}\hfill
         \includegraphics[width=0.15\textwidth,trim={40  100 40 30},clip]{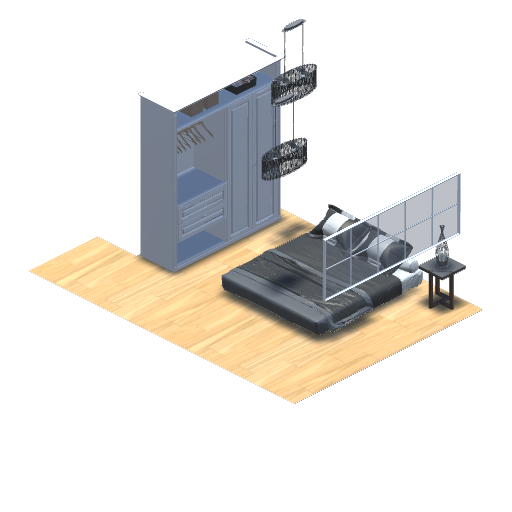}\hfill
         \label{fig:res_v1}
     \end{subfigure}
     \begin{subfigure}[t]{\figw\textwidth}
         \centering
         \caption*{Results Ours (Loss-only): Bedrooms}
         \includegraphics[width=0.15\textwidth,trim={0 80 0 80},clip]{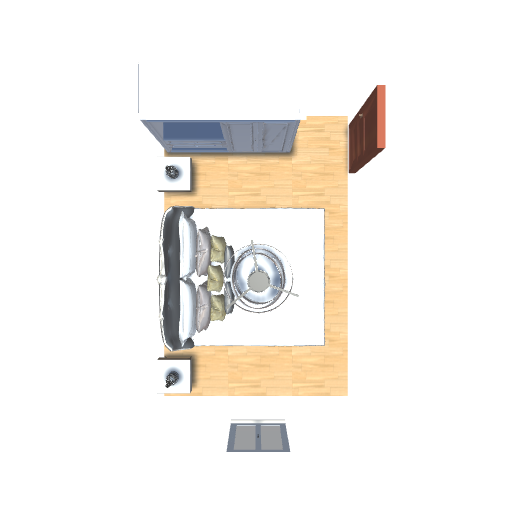} \hfill
         \includegraphics[width=0.15\textwidth,trim={0 80 0 80},clip]{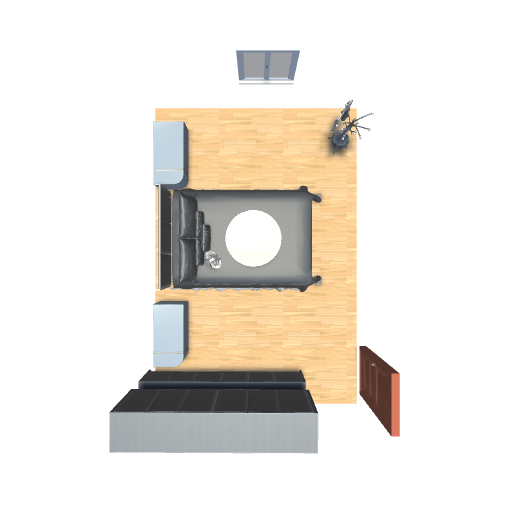}\hfill
         \includegraphics[width=0.15\textwidth,trim={0 80 0 80},clip]{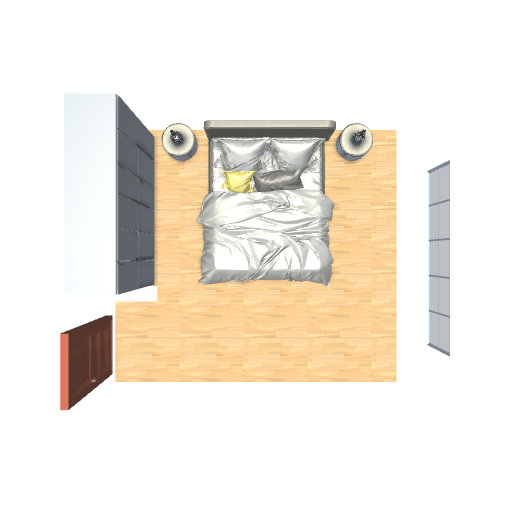}\hfill
         \includegraphics[width=0.15\textwidth,trim={0 80 0 80},clip]{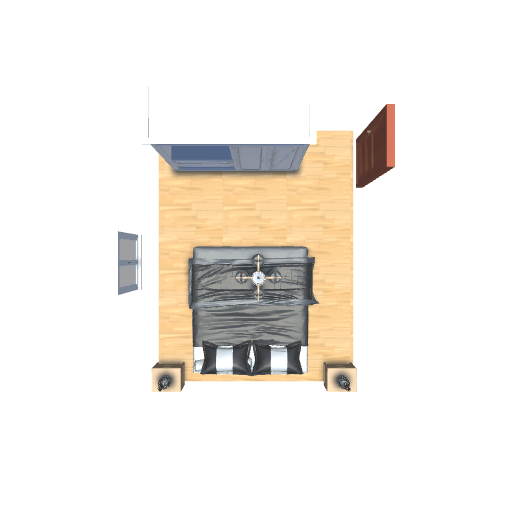}\hfill
         \includegraphics[width=0.15\textwidth,trim={0 80 0 80},clip]{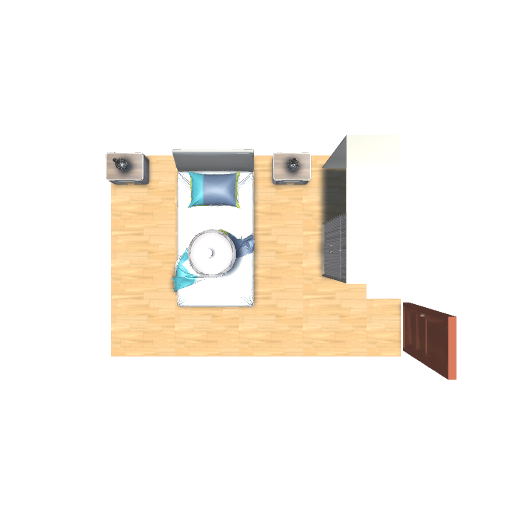}\hfill
         \includegraphics[width=0.15\textwidth,trim={0 80 0 80},clip]{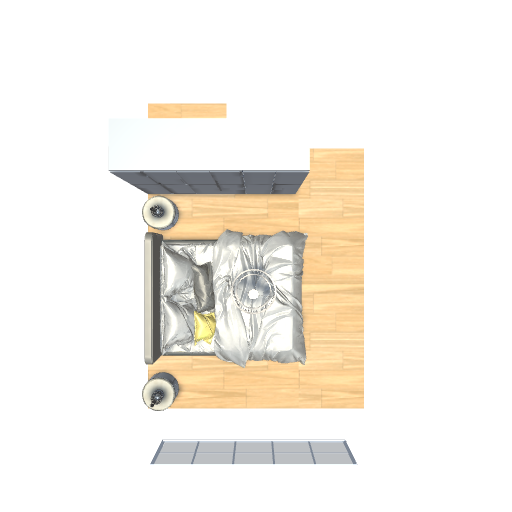}\hfill
         \includegraphics[width=0.15\textwidth,trim={40  100 40 30},clip]{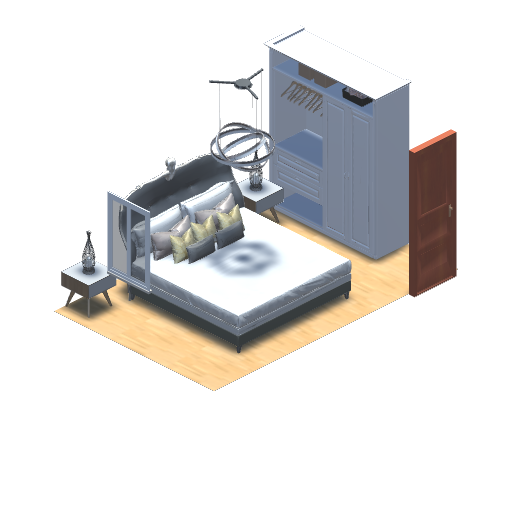} \hfill
         \includegraphics[width=0.15\textwidth,trim={40  100 40 30},clip]{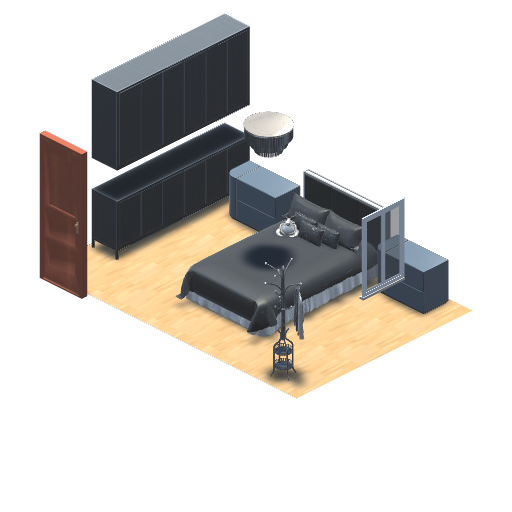}\hfill
         \includegraphics[width=0.15\textwidth,trim={40  100 40 30},clip]{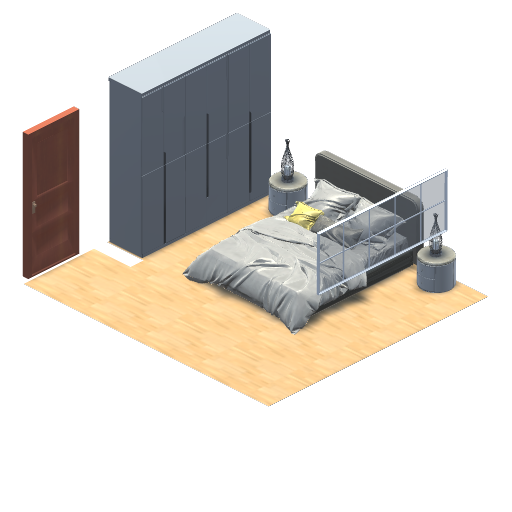}\hfill
         \includegraphics[width=0.15\textwidth,trim={40  100 40 30},clip]{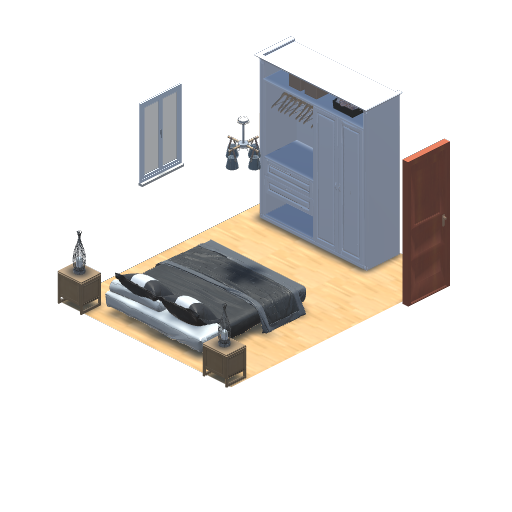}\hfill
         \includegraphics[width=0.15\textwidth,trim={40  100 40 30},clip]{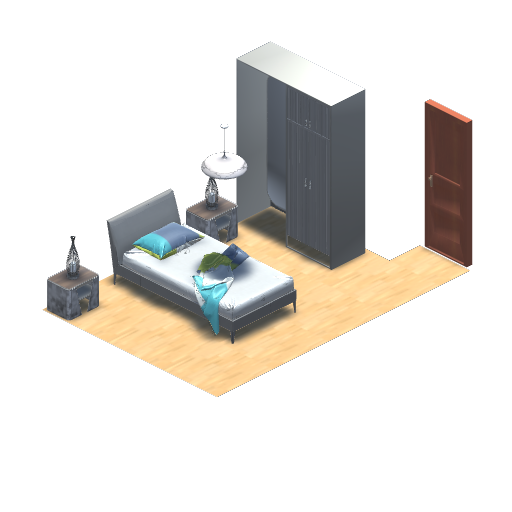}\hfill
         \includegraphics[width=0.15\textwidth,trim={40  100 40 30},clip]{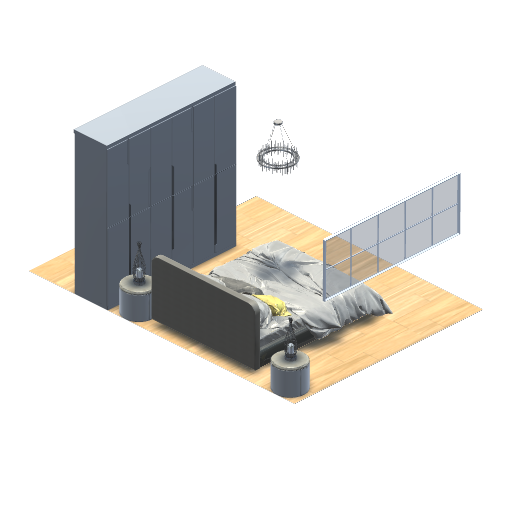}\hfill
         \label{fig:res_v2}
     \end{subfigure}
        \caption{Room-conditioned synthesis results for our model and its ablations.}
        \label{fig:res:comp:gt_at_v3}
\end{figure*}

\section{Source Code}
The author's source code is available at \url{https://github.com/kleimerTU/HumanCentricLayouts}

\end{document}